\documentclass[twocolumn]{aastex62}

\submitjournal{ApJ}
\usepackage{lineno}

\shorttitle{Gru II Chemical Analysis}

\begin{document}
  
\title{Chemical Analysis of the Ultra-Faint Dwarf Galaxy Grus~II.\\ 
Signature of high-mass stellar nucleosynthesis \footnote{This paper includes data gathered with the 6.5m Magellan Telescopes located at Las Campanas Observatory, Chile.}}


\author[0000-0001-6154-8983]{T.~T.~Hansen}
\altaffiliation{Mitchell Astronomy Fellow}
\affil{Mitchell Institute for Fundamental Physics and Astronomy and Department of Physics and Astronomy, Texas A\&M University, College Station, TX 77843-4242, USA}

\author{J.~L.~Marshall}
\affil{Mitchell Institute for Fundamental Physics and Astronomy and Department of Physics and Astronomy, Texas A\&M University, College Station, TX 77843-4242, USA}

\author{J.~D.~Simon}
\affil{Carnegie Institution for Science, 813 Santa Barbara St., Pasadena, CA 91101, USA}

\author{T.~S.~Li}
\altaffiliation{NHFP Einstein Fellow}
\affil{Carnegie Institution for Science, 813 Santa Barbara St., Pasadena, CA 91101, USA}
\affil{Department of Astrophysical Sciences, Princeton University, Peyton Hall, Princeton, NJ 08544, USA}
\affil{Fermi National Accelerator Laboratory, P. O. Box 500, Batavia, IL 60510, USA}
\affil{Kavli Institute for Cosmological Physics, University of Chicago, Chicago, IL 60637, USA}

\author{R.~A.~Bernstein} 
\affil{Carnegie Institution for Science, 813 Santa Barbara St., Pasadena, CA 91101, USA}

\author{A.~B.~Pace}
\affil{McWilliams Center for Cosmology, Carnegie Mellon University, 5000
Forbes Ave, Pittsburgh, PA 15213, USA}

\author{P.~Ferguson}
\affil{Mitchell Institute for Fundamental Physics and Astronomy and Department of Physics and Astronomy, Texas A\&M University, College Station, TX 77843-4242, USA}

\author{D.~Q.~Nagasawa}
\affil{Mitchell Institute for Fundamental Physics and Astronomy and Department of Physics and Astronomy, Texas A\&M University, College Station, TX 77843-4242, USA}

\author{K.~Kuehn}
\affil{Australian Astronomical Optics, Macquarie University, North Ryde, NSW 2113, Australia}
\affil{Lowell Observatory, 1400 W Mars Hill Rd, Flagstaff,  AZ 86001, USA}

\author{D.~Carollo}
\affil{INAF- Osservatorio Astronomico di Torino - Strada Osservatorio 20, Pino Torinese, 10020, Italy}

\author{M.~Geha}
\affil{Department of Astronomy, Yale University, New Haven, CT 06520, USA}

\author{D.~James}
\affil{Harvard-Smithsonian Center for Astrophysics, Cambridge, MA 02138, USA}

\author{A.~Walker}
\affil{Cerro Tololo Inter-American Observatory, National Optical Astronomy Observatory, Casilla 603, La Serena, Chile}

\author{H.~T.~Diehl}
\affil{Fermi National Accelerator Laboratory, P. O. Box 500, Batavia, IL 60510, USA}

\author{M.~Aguena}
\affil{Departamento de F\'isica Matem\'atica, Instituto de F\'isica, Universidade de S\~ao Paulo, CP 66318, S\~ao Paulo, SP, 05314-970, Brazil}
\affil{Laborat\'orio Interinstitucional de e-Astronomia - LIneA, Rua Gal. Jos\'e Cristino 77, Rio de Janeiro, RJ - 20921-400, Brazil}

\author{S.~Allam}
\affil{Fermi National Accelerator Laboratory, P. O. Box 500, Batavia, IL 60510, USA}
\affil{School of Mathematics and Physics, University of Queensland,  Brisbane, QLD 4072, Australia}

\author{S.~Avila}
\affil{Instituto de Fisica Teorica UAM/CSIC, Universidad Autonoma de Madrid, 28049 Madrid, Spain}

\author{E.~Bertin}
\affil{CNRS, UMR 7095, Institut d'Astrophysique de Paris, F-75014, Paris, France}
\affil{Sorbonne Universit\'es, UPMC Univ Paris 06, UMR 7095, Institut d'Astrophysique de Paris, F-75014, Paris, France}

\author{D.~Brooks}
\affil{Department of Physics \& Astronomy, University College London, Gower Street, London, WC1E 6BT, UK}

\author{E.~Buckley-Geer}
\affil{Fermi National Accelerator Laboratory, P. O. Box 500, Batavia, IL 60510, USA}

\author{D.~L.~Burke}
\affil{Kavli Institute for Particle Astrophysics \& Cosmology, P. O. Box 2450, Stanford University, Stanford, CA 94305, USA}
\affil{SLAC National Accelerator Laboratory, Menlo Park, CA 94025, USA}

\author{A.~Carnero~Rosell}
\affil{Centro de Investigaciones Energ\'eticas, Medioambientales y Tecnol\'ogicas (CIEMAT), Madrid, Spain}
\affil{Laborat\'orio Interinstitucional de e-Astronomia - LIneA, Rua Gal. Jos\'e Cristino 77, Rio de Janeiro, RJ - 20921-400, Brazil}

\author{M.~Carrasco~Kind}
\affil{Department of Astronomy, University of Illinois at Urbana-Champaign, 1002 W. Green Street, Urbana, IL 61801, USA}
\affil{National Center for Supercomputing Applications, 1205 West Clark St., Urbana, IL 61801, USA}

\author{J.~Carretero}
\affil{Institut de F\'{\i}sica d'Altes Energies (IFAE), The Barcelona Institute of Science and Technology, Campus UAB, 08193 Bellaterra (Barcelona) Spain}

\author{M.~Costanzi}
\affil{INAF-Osservatorio Astronomico di Trieste, via G. B. Tiepolo 11, I-34143 Trieste, Italy}
\affil{Institute for Fundamental Physics of the Universe, Via Beirut 2, 34014 Trieste, Italy}

\author{L.~N.~da Costa}
\affil{Laborat\'orio Interinstitucional de e-Astronomia - LIneA, Rua Gal. Jos\'e Cristino 77, Rio de Janeiro, RJ - 20921-400, Brazil}
\affil{Observat\'orio Nacional, Rua Gal. Jos\'e Cristino 77, Rio de Janeiro, RJ - 20921-400, Brazil}

\author{S.~Desai}
\affil{Department of Physics, IIT Hyderabad, Kandi, Telangana 502285, India}

\author{J.~De~Vicente}
\affil{Centro de Investigaciones Energ\'eticas, Medioambientales y Tecnol\'ogicas (CIEMAT), Madrid, Spain}

\author{P.~Doel}
\affil{Department of Physics \& Astronomy, University College London, Gower Street, London, WC1E 6BT, UK}

\author{K.~Eckert}
\affil{Department of Physics and Astronomy, University of Pennsylvania, Philadelphia, PA 19104, USA}

\author{T.~F.~Eifler}
\affil{Department of Astronomy/Steward Observatory, University of Arizona, 933 North Cherry Avenue, Tucson, AZ 85721-0065, USA}
\affil{Jet Propulsion Laboratory, California Institute of Technology, 4800 Oak Grove Dr., Pasadena, CA 91109, USA}

\author{S.~Everett}
\affil{Santa Cruz Institute for Particle Physics, Santa Cruz, CA 95064, USA}

\author{I.~Ferrero}
\affil{Institute of Theoretical Astrophysics, University of Oslo, P.O. Box 1029 Blindern, NO-0315 Oslo, Norway}

\author{J.~Frieman}
\affil{Fermi National Accelerator Laboratory, P. O. Box 500, Batavia, IL 60510, USA}
\affil{Kavli Institute for Cosmological Physics, University of Chicago, Chicago, IL 60637, USA}

\author{J.~Garc\'ia-Bellido}
\affil{Instituto de Fisica Teorica UAM/CSIC, Universidad Autonoma de Madrid, 28049 Madrid, Spain}

\author{E.~Gaztanaga}
\affil{Institut d'Estudis Espacials de Catalunya (IEEC), 08034 Barcelona, Spain}
\affil{Institute of Space Sciences (ICE, CSIC),  Campus UAB, Carrer de Can Magrans, s/n,  08193 Barcelona, Spain}

\author{D.~W.~Gerdes}
\affil{Department of Astronomy, University of Michigan, Ann Arbor, MI 48109, USA}
\affil{Department of Physics, University of Michigan, Ann Arbor, MI 48109, USA}

\author{D.~Gruen}
\affil{Department of Physics, Stanford University, 382 Via Pueblo Mall, Stanford, CA 94305, USA}
\affil{Kavli Institute for Particle Astrophysics \& Cosmology, P. O. Box 2450, Stanford University, Stanford, CA 94305, USA}
\affil{SLAC National Accelerator Laboratory, Menlo Park, CA 94025, USA}

\author{R.~A.~Gruendl}
\affil{Department of Astronomy, University of Illinois at Urbana-Champaign, 1002 W. Green Street, Urbana, IL 61801, USA}
\affil{National Center for Supercomputing Applications, 1205 West Clark St., Urbana, IL 61801, USA}

\author{J.~Gschwend}
\affil{Laborat\'orio Interinstitucional de e-Astronomia - LIneA, Rua Gal. Jos\'e Cristino 77, Rio de Janeiro, RJ - 20921-400, Brazil}
\affil{Observat\'orio Nacional, Rua Gal. Jos\'e Cristino 77, Rio de Janeiro, RJ - 20921-400, Brazil}

\author{G.~Gutierrez}
\affil{Fermi National Accelerator Laboratory, P. O. Box 500, Batavia, IL 60510, USA}

\author{S.~R.~Hinton}
\affil{School of Mathematics and Physics, University of Queensland,  Brisbane, QLD 4072, Australia}

\author{D.~L.~Hollowood}
\affil{Santa Cruz Institute for Particle Physics, Santa Cruz, CA 95064, USA}

\author{K.~Honscheid}
\affil{Center for Cosmology and Astro-Particle Physics, The Ohio State University, Columbus, OH 43210, USA}
\affil{Department of Physics, The Ohio State University, Columbus, OH 43210, USA}

\author{N.~Kuropatkin}
\affil{Fermi National Accelerator Laboratory, P. O. Box 500, Batavia, IL 60510, USA}
\affil{Department of Physics \& Astronomy, University College London, Gower Street, London, WC1E 6BT, UK}

\author{M.~A.~G.~Maia}
\affil{Laborat\'orio Interinstitucional de e-Astronomia - LIneA, Rua Gal. Jos\'e Cristino 77, Rio de Janeiro, RJ - 20921-400, Brazil}
\affil{Observat\'orio Nacional, Rua Gal. Jos\'e Cristino 77, Rio de Janeiro, RJ - 20921-400, Brazil}

\author{M.~March}
\affil{Department of Physics and Astronomy, University of Pennsylvania, Philadelphia, PA 19104, USA}

\author{R.~Miquel}
\affil{Instituci\'o Catalana de Recerca i Estudis Avan\c{c}ats, E-08010 Barcelona, Spain}
\affil{Institut de F\'{\i}sica d'Altes Energies (IFAE), The Barcelona Institute of Science and Technology, Campus UAB, 08193 Bellaterra (Barcelona) Spain}

\author{A.~Palmese}
\affil{Fermi National Accelerator Laboratory, P. O. Box 500, Batavia, IL 60510, USA}
\affil{Kavli Institute for Cosmological Physics, University of Chicago, Chicago, IL 60637, USA}

\author{F.~Paz-Chinch\'{o}n}
\affil{Department of Astronomy, University of Illinois at Urbana-Champaign, 1002 W. Green Street, Urbana, IL 61801, USA}
\affil{National Center for Supercomputing Applications, 1205 West Clark St., Urbana, IL 61801, USA}

\author{A.~A.~Plazas}
\affil{Department of Astrophysical Sciences, Princeton University, Peyton Hall, Princeton, NJ
08544, USA}

\author{E.~Sanchez}
\affil{Centro de Investigaciones Energ\'eticas, Medioambientales y Tecnol\'ogicas (CIEMAT), Madrid, Spain}

\author{B.~Santiago}
\affil{Instituto de F\'\i sica, UFRGS, Caixa Postal 15051, Porto Alegre, RS - 91501-970, Brazil}
\affil{Laborat\'orio Interinstitucional de e-Astronomia - LIneA, Rua Gal. Jos\'e
Cristino 77, Rio de Janeiro, RJ - 20921-400, Brazil}

\author{V.~Scarpine}
\affil{Fermi National Accelerator Laboratory, P. O. Box 500, Batavia, IL 60510, USA}

\author{S.~Serrano}
\affil{Institut d'Estudis Espacials de Catalunya (IEEC), 08034 Barcelona, Spain}
\affil{Institute of Space Sciences (ICE, CSIC),  Campus UAB, Carrer de Can Magrans, s/n,  08193 Barcelona, Spain}

\author{M.~Smith}
\affil{School of Physics and Astronomy, University of Southampton,  Southampton, SO17 1BJ, UK}

\author{M.~Soares-Santos}
\affil{Brandeis University, Physics Department, 415 South Street, Waltham MA 02453}

\author{E.~Suchyta}
\affil{Computer Science and Mathematics Division, Oak Ridge National Laboratory, Oak Ridge, TN 37831}

\author{M.~E.~C.~Swanson}
\affil{National Center for Supercomputing Applications, 1205 West Clark St., Urbana, IL 61801, USA}

\author{G.~Tarle}
\affil{Department of Physics, University of Michigan, Ann Arbor, MI 48109, USA}

\author{T.~N.~Varga}
\affil{Max Planck Institute for Extraterrestrial Physics, Giessenbachstrasse, 85748 Garching, Germany}
\affil{Universit{\"a}ts-Sternwarte, Fakult{\"a}t f{\"u}r Physik, Ludwig-Maximilians Universit{\"a}t M{\"u}nchen, Scheinerstr. 1, 81679 M{\"u}nchen, Germany}

\author{R.~Wilkinson}
\affil{Department of Physics and Astronomy, Pevensey Building, University of Sussex, Brighton, BN1 9QH, UK}

\collaboration{(DES Collaboration)}

\correspondingauthor{T. T. Hansen}
\email{thansen@tamu.edu}


\begin{abstract}
We present a detailed abundance analysis of the three brightest member stars at the top of the giant branch of the ultra faint dwarf galaxy Grus~II. All stars exhibit a higher than expected $\mathrm{[Mg/Ca]}$ ratio compared to metal-poor stars in other ultra faint dwarf galaxies and in the Milky Way halo. Nucleosynthesis in high mass ($\geqslant 20$M$_\odot$) core-collapse supernovae has been shown to create this signature. The abundances of this small sample (3) stars suggests the chemical enrichment of Grus~II could have occurred through substantial high-mass stellar evolution, and is consistent with the framework of a top-heavy initial mass function. However, with only three stars it can not be ruled out that the abundance pattern is the result of a stochastic chemical enrichment at early times in the galaxy. The most metal-rich of the three stars also possesses a small enhancement in rapid neutron-capture ($r$-process) elements. The abundance pattern of the $r$-process elements in this star matches the scaled $r$-process pattern of the solar system and $r$-process enhanced stars in other dwarf galaxies and in the Milky Way halo, hinting at a common origin for these elements across a range of environments. All current proposed astrophysical sites of $r$-process element production are associated with high-mass stars, thus the possible top-heavy initial mass function of Grus~II would increase the likelihood of any of these events occurring. The time delay between the $\alpha$ and $r$-process element enrichment of the galaxy favours a neutron star merger as the origin of the $r$-process elements in Grus~II.
\end{abstract}

\section{Introduction \label{sec:intro}}
Recent exploration of the chemical abundances in ultra faint dwarf (UFD) galaxies have revealed abundance patterns of these objects similar to those which have been detected in the majority of metal-poor Milky Way (MW) halo stars \citep{frebel2015}. However, some notable outliers have also been observed, for example Reticulum~II (Ret~II) and Tucana~III (Tuc~III) both show enhancement in rapid neutron-capture ($r$-process) elements \citep{ji2016a,roederer2016,hansen2017,marshall2019}. This chemical signature is also seen in a very small fraction of MW halo stars \citep{barklem2005,hansen2018}. More rare abundance patterns have also been detected in the UFD galaxies Horologium~I (Hor~I) and Hercules (Her), with Hor~I exhibiting a deficiency in $\alpha$ elements like Mg and Ca \citep{nagasawa2018}, and some stars in Her showing very high $\mathrm{[Mg/Ca]}$ ratios compared to the average metal-poor MW halo and UFD galaxy stars \citep{koch2008}. 

The UFD galaxies are small dark matter dominated systems which have slower chemical enrichment compared to the larger dwarf spheroidal galaxies and the MW \citep{simon2019}. Hence the chemical abundances of the stars in these systems provide a window to study single nucleosynthesis events in closed environments. The peculiar chemical patterns described above are particularly important to study rare chemical enrichment events, such as neutron star mergers (NSMs) or special types of core-collapse supernovae (CCSNe) such as magnetorotational supernovae \citep{winteler2012}, collapsars \citep{siegel2019}, and supernovae with mixing and fallback \citep{umeda2003}. 

The chemistry of the first generation of low mass stars to form in a galaxy is tightly linked to the mass range of the first population of massive stars (Pop~III) that formed and terminally evolved in them. \citet{placco2016} used the abundance pattern of the most metal-poor MW halo stars along with nucleosynthesis models from \citet{heger2010} to constrain the mass range of the Pop~III stars responsible for the initial chemical enrichment of the Galaxy. Similar analysis in UFD galaxies are hampered by the small number of stars for which detailed chemical abundances can be determined in these systems. 

On the other hand, UFD galaxies have been used to study the effect of environment on the initial stellar mass function (IMF). The long relaxation times of this type of galaxy means that the low mass stellar IMF can be measured directly from the number of stars below the main-sequence turn-off present in the galaxy today \citep{geha2013}. The low luminosity of these systems, however, makes it difficult to obtain the photometric observations needed for this measurement. 
For those UFD galaxies where the measurements have been made (Boo~I, CVn~II, ComBer, Her, Leo~IV, UMa~I), evidence of variations of the IMF with environment have been found. For example \citet{gennaro2018} measured the IMF for a sample of six UFDs and found that their sub-solar stellar IMFs were generally more bottom-light than the IMF for MW disk stars, though with a large scatter within the sample. In another study \citet{geha2013} measured the IMF for Hercules and Leo IV and also found bottom light IMFs for these galaxies. For other types of galaxies, some correlations have been detected between IMFs and galaxy properties, with the largest galaxies having bottom-heavy IMFs and smaller galaxies having more bottom-light IMFs \citep{geha2013}. This variation is also supported by the results from \citet{kalirai2013}, who measured the IMF of the Small Magellanic Cloud and found a shallower slope than what is determined in the MW. 

It is important to note that even though the low mass stellar populations of today's UFD galaxies provides a direct way to measure the IMF of these systems, this does not provide any information on the IMF of the higher stellar mass population previously present in the galaxy. However, here the chemistry of the metal-poor stars may provide some clues. The chemical abundances of the most metal-poor stars in the UFD galaxies constitute a record of the nucleosynthesis happening in the first generation of massive stars formed in these systems. Hence, the chemical abundance pattern can provide constraints on the mass range of the first stars to form in the UFD galaxy. 

In this paper we analyse the UFD galaxy Grus~II (Gru~II). Gru~II was identified as a candidate MW satellite galaxy in the Dark Energy Survey (DES) at a distance of 53~kpc \citep{drlicawagner2015}. Spectroscopic follow-up of the system was presented by \citet{simon2020}, who was unable to resolve a velocity dispersion or detect a metallicity spread, indicative of dwarf galaxies, for the system. However, based on the large physical size ($r_{1/2} = 94$~pc) and low metallicity ($\mathrm{[Fe/H]} = -2.51 \pm 0.11$), \citet{simon2020} classify the system as a likely dwarf galaxy. We present here a detailed chemical abundance analysis of the three brightest member stars at the top of the giant branch of Gru~II. The paper is organized as follows: Observations and analysis of the stars are described in Sections \ref{sec:obs} and \ref{sec:param}, Section \ref{sec:results} presents our results which are discussed in Section \ref{sec:discus}, and Section \ref{sec:sum} provides a summary.   

\begin{deluxetable*}{lcccccccccccc}[hbt!]
\rotate
\tablecaption{Observing log\label{tab:obslog}} 
\tablewidth{0pt}
\tablehead{Object Name & R.A. & Dec.& $g$& $r$ & $i$ & $z$ &Date & $t_{exp}$ & $\rm{SNR}$ &$\rm{SNR}$ &$V_{hel}$ & Member \\
&(J2000)&(J2000)&(mag)&(mag)&(mag)&(mag)&(MJD)&  (sec) & @4100\AA\ & @5500\AA\ &km~s$^{-1}$}
\startdata
DES J220423.91-463702.5 & 22:04:23.90 & $-$46:37:02.48 & 17.524 &16.741 & 16.456&16.289& 57605 &7x1800 & 12 & 32 &$-$106.9$\pm$0.6 & Yes\\ 
DES J220409.98-462102.2 & 22:04:09.98 & $-$46:21:02.21 & 18.157 &17.477 & 17.227&17.086& 57607 &3x3000 & 9 & 17 &$-$106.6$\pm$0.6 & Yes\\ 
DES J220352.01-462446.5 & 22:03:51.90 & $-$46:24:46.40 & 19.396 &18.860 & 18.678&18.568& 57983 &2x1800 & 9$^*$ & 12$^*$ &$-$106.5$\pm$1.1 & Yes\\ 
                        &             &                &        & & && 58430 &3x1800+824&9$^*$&12$^*$&$-$106.4$\pm$0.7\\
DES J220318.62-464116.4 & 22:03:18.62 & $-$46:41:16.44 & 18.726 & & && 58433 &600+1800 & 5 & 7 &$+$76.3 $\pm$0.7 & No\\ 
DES J220253.88-463522.6 & 22:02:53.88 & $-$46:35:22.56 & 18.703 & & && 58341 &1800 & 3 & 5 &$+$40.0$\pm$0.9 & No\\ 
\enddata
\tablecomments{$^*$ Signal-to-noise ratio (SNR) of the combined spectrum per pixel.}
\end{deluxetable*}

\section{Observations \label{sec:obs}}
\begin{figure}
\centering 
\includegraphics[scale=0.5]{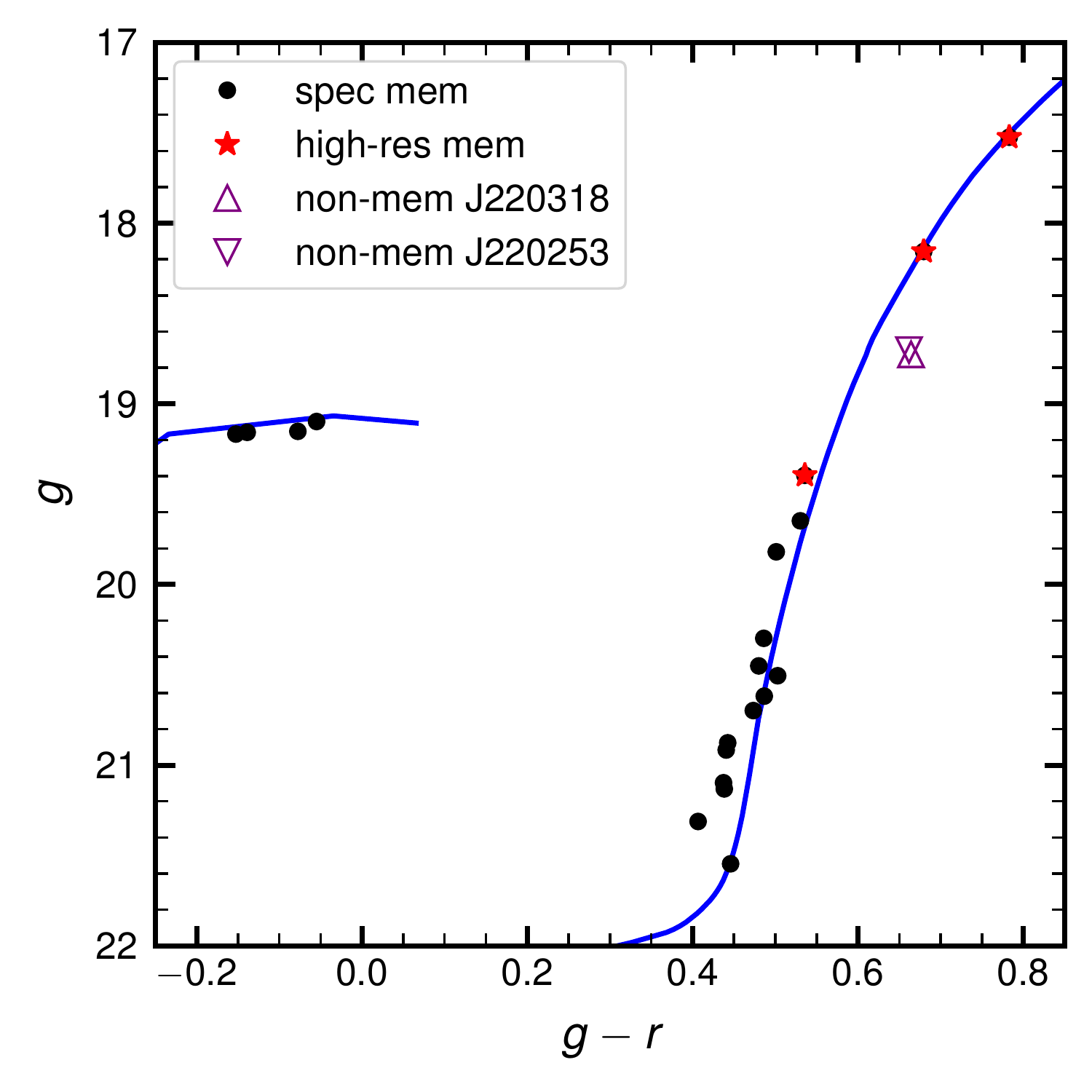}
\caption{Color-magnitude diagram for Gru~II. Black dots are confirmed spectroscopic members from \citet{simon2020}. Open triangle symbols are observed non members; filled star symbols are observed members. Blue curves show a Dartmouth isochrone \citep{dotter2008} with $\mathrm{[Fe/H]} = -2.2$ and age = 12.5 Gyr, and a M92 blue horizontal branch ridgeline from \citet{belokurov2007} transformed to the DES photometric system and shifted to the distance of Gru~II. \label{fig:cmd}
}
\end{figure}

A total sample of five stars were observed with the MIKE spectrograph \citep{bernstein2003} at the Magellan-Clay Telescope at Las Campanas Observatory. Table \ref{tab:obslog} lists the targets, observing dates, and exposure times. The two brightest stars DES~J220423.91-463702.5 and DES~J220409.98-462102.2 (hereafter referred to as J220423 and J220409) were observed during a run in August 2016. The first spectrum of DES J220352.01-462446.5 (J220352) was observed in August 2017 and the second in November 2018 along with spectra of DES~J220318.62-464116.4 (J220318) and DES~J220253.88-463522.6 (J220253). Three of the stars (J220423, J220409, and J220352) were selected as spectroscopically confirmed members from medium-resolution velocity and metallicity measurements \citep{simon2020}. The brightest star, J220423, was first identified as a member from an observing run with the Anglo-Australia Telescope (AAT) in July 2016. While the AAT observing run was mostly for targeting the Tucana III stream \citep[see details in ][]{li2018}, a small fraction of time was also spent on bright candidate members in Gru~II and Tucana IV during nights with cirrus cloud coverage, to identify bright member stars for MIKE observations. The membership of J220423 was subsequently confirmed with Magellan/IMACS spectroscopy \citep{simon2020}. The remaining two stars, J220318 and J220253, were selected as member candidates from \citet{pace2019}, although with relatively low membership probabilities ($P_i = 0.47$ and $P_i = 0.14$, respectively). Radial velocities derived from short MIKE exposures establish these as non-members (see Table \ref{tab:obslog}). Figure \ref{fig:cmd} shows a color-magnitude diagram of Gru~II member stars; star symbols mark the five stars observed, filled symbols are members while open symbols are not members.

\begin{figure*}
\centering
\includegraphics[scale=0.45]{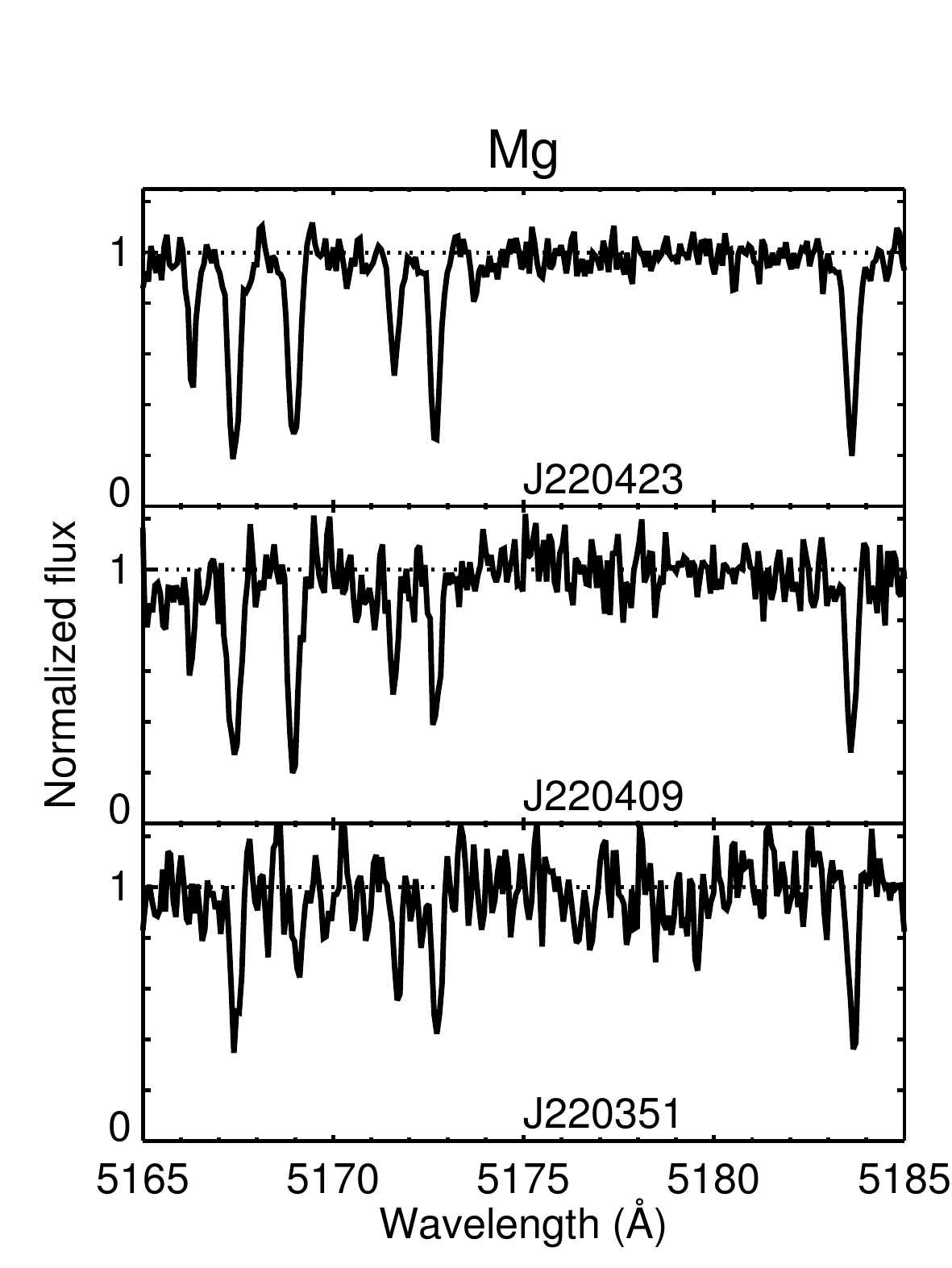}
\includegraphics[scale=0.45]{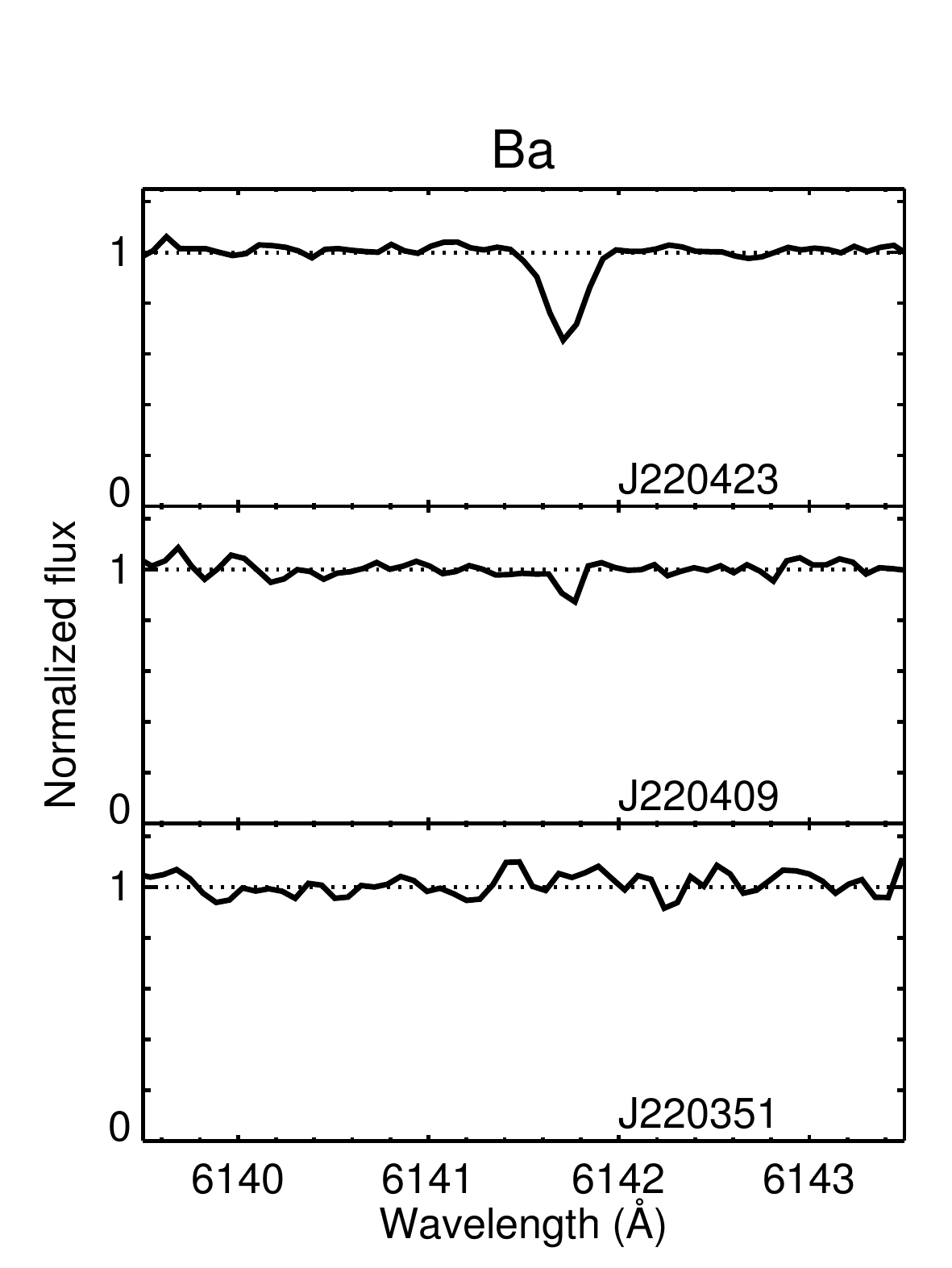}
\includegraphics[scale=0.45]{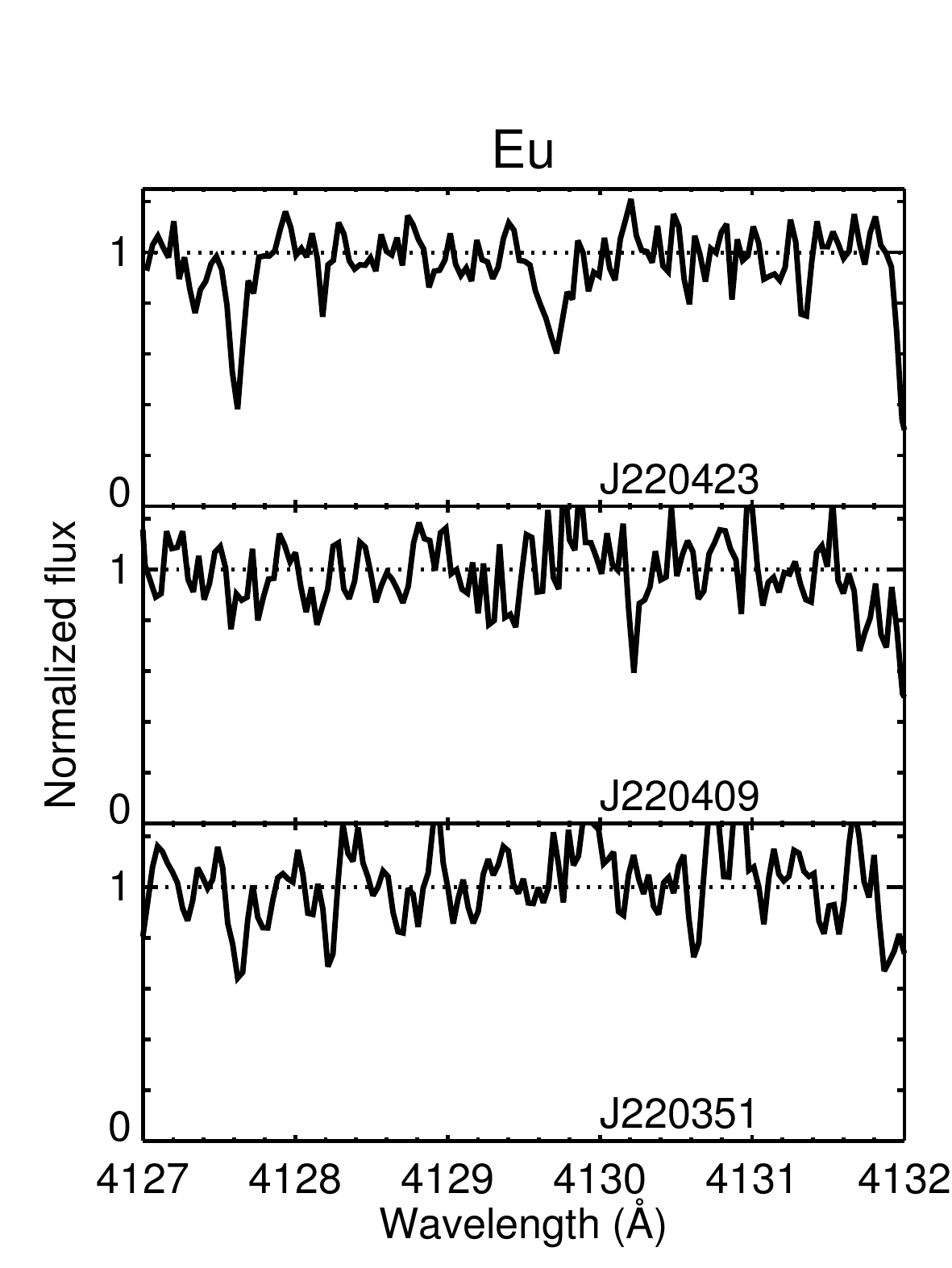}
\caption{Spectra of the three stars around the \ion{Mg}{1} b lines (left), the \ion{Ba}{2} 6141{\AA} line (middle), and the \ion{Eu}{2} 4129{\AA} line (right). \label{fig:spec}}
\end{figure*}

The spectra of the stars were obtained using a 0.7 arcsec slit with 2x2 pixel binning resulting in a spectral resolution of R=$\lambda / \Delta\lambda\approx$ 41,000 in the blue and 32,000 in the red. The spectra cover 3310\AA$<\lambda<$5000\AA\ in the blue channel and 4830\AA$<\lambda<$9160\AA\ in the red. The data were reduced using the latest version of the CarPy MIKE pipeline \citep{kelson2000,kelson2003}.  For stars with multiple spectra taken on the same run (J220409, J220423, and J220318) these were co-added (summed) during reduction while the spectra from multiple runs (J220352) were co-added after reduction. Following reduction the spectra were normalized and shifted to rest wavelength. Final reduced and normalized spectra around the \ion{Mg}{1} b lines, the \ion{Ba}{2} 6141{\AA} line, and the \ion{Eu}{2} 4129{\AA} line are shown in Figure \ref{fig:spec}. Signal to noise ratios (SNR) per pixel of the final reduced spectra at 4100{\AA}~ and 5500{\AA} are listed in Table \ref{tab:obslog}. Heliocentric radial velocities of the stars were determined by cross-correlation \citep{tonry1979} of the target spectra with a spectrum of the radial-velocity standard star HD122563 ($V_{hel}=-26.51$~km~s$^{-1}$; \citet{chubak2012}) obtained with the same instrument setup as the target stars. Depending on the SNR of the target spectra, 30 to 50 individual orders were used for the cross-correlation. The mean value of the resulting velocities along with the standard deviation are listed in Table \ref{tab:obslog}. Our values are different from the velocities determined for the stars by \citet{simon2020}, who find $V_{hel}= -108.6\pm1.0$~km~s$^{-1}$, $-108.1\pm1.0$~km~s$^{-1}$, and  $-109.3\pm1.0$~km~s$^{-1}$ for J220423, J220409, and J220352, respectively. However, J220423 and J220409 were both observed at high airmass, 1.8 and 1.6 respectively, which can cause offsets in the radial velocities \citep{ji2020}, and  J220352 is likely in a binary system \citep{simon2020}.

\begin{deluxetable*}{lccccc}
\caption{Measured Stellar Parameters\label{tab:params}}
\tablehead{ID & $T_{eff,photo}$ & $T_{eff,spec}^*$& log $g$ & $\xi$  & $\mathrm{[Fe/H]}$ \\
&(K)&(K)&(cgs)&(km~s$^{-1}$)&(dex)}
\startdata
DES J220423 & 4556$\pm 97$ &$4585 \pm 150$ & $1.22\pm 0.3$ & $2.25\pm 0.3$ & $-2.49\pm0.18$ \\  
DES J220409 & 4740$\pm 73$ &$4720 \pm 150$ & $1.55\pm 0.3$ & $2.35\pm 0.3$ & $-2.69\pm0.17$ \\ 
DES J220352 & 5121$\pm 81$ &$4920 \pm 150$ & $1.91\pm 0.3$ & $2.25\pm 0.3$ & $-2.94\pm0.15$\\ 
\enddata
\tablecomments{$^*$ Used to determine log $g$, $\xi$, $\mathrm{[Fe/H]}$, and for abundance analysis.}
\end{deluxetable*}

\section{Stellar Parameter Determination and Abundance Analysis \label{sec:param}}
Stellar parameter determination and abundance analysis was done following the techniques described in \citet{hansen2017} and \citet{marshall2019}, using the 2017 version of MOOG \citep{sneden1973} and making the assumption of local thermodynamic equilibrium (LTE) and including Rayleigh scattering treatment as described by \citet{sobeck2011}\footnote{https://github.com/alexji/moog17scat}.
The stellar parameters for the three stars were determined spectroscopically from equivalent width (EW) measurements of \ion{Fe}{1} and \ion{Fe}{2} lines. We measure the EW by fitting Gaussian profiles to the absorption lines in the continuum-normalized spectra. Uncertainties on the EWs were computed using $\sigma_{EW}=1.5\sqrt{FWHM*\delta x}/\rm{SNR}$ from \citet{cayrel1988}, where SNR is the signal to noise per pixel and $\delta x$ is the pixel size. First estimates of effective temperatures were determined from excitation equilibrium of \ion{Fe}{1} lines. These were then placed on a photometric scale using the relation from \cite{frebel2013}. Using the corrected temperatures, surface gravities ($\log g$) were determined from ionization equilibrium between the \ion{Fe}{1} and \ion{Fe}{2} lines. Finally microturbulent velocities ($\xi$) were determined by removing any trend in line abundances with reduced EW for the \ion{Fe}{1} lines. Final stellar parameters along with estimated uncertainties are presented in Table \ref{tab:params} and lines used for the parameter determination of each star are listed in Table \ref{tab:ew}. All three stars are giants. For comparison Table \ref{tab:params} also lists the photometric temperatures for our stars which are in good agreement with the corrected spectroscopic temperatures. The photometric temperatures were derived by converting the $g,r,i,$ and $z$ colors listed in Table \ref{tab:obslog} from DES DR1\citep{abbott2018} to the corresponding $B,V,R,$ and $I$ colors \citep[][R. Lupton 2005\footnote{http://www.sdss3.org/dr8/algorithms/sdssUBVRITransform.php}]{drlicawagner2018}, and using the $B-V$, $V-R$, $R-I$, and $V-I$ color temperature relations from \citet{casagrande2010}. Listed in Table \ref{tab:params} is the average photometric temperature and the standard deviation. Our derived metallicities are in good agreement with the metallicities derived from the Calcium Triplet lines by \citet{simon2020}, who find $\mathrm{[Fe/H]}= -2.62\pm0.16$,  $-2.72\pm0.16$, and $-2.93\pm0.22$ for J220423, J220409, and J220352, respectively. 

Abundances were derived from EW measurements and spectral synthesis. EWs are used for strong non-blended lines while spectral synthesis is used for weaker and or blended lines and for lines affected by isotopic and or hyperfine splitting. Measured EWs used for abundance determination are given in Table \ref{tab:abunew}. All abundances were derived using $\alpha$-enhanced ($\mathrm{[\alpha/Fe]} = +0.4$) 1D LTE ATLAS9 model atmospheres \citep{castelli2003} and the solar photosphere abundances from \cite{asplund2009}. As can be seen in Table \ref{tab:abun}, our stars are not enhanced in all $\alpha$ elements, hence using non-$\alpha$-enhanced atmosphere models might be a better fit for our stars. However, \citet{nagasawa2018} showed that for low SNR spectra the difference in abundances was not more than 0.05 dex between using non-$\alpha$-enhanced and $\alpha$-enhanced atmosphere models to derive abundances for their low $\alpha$ stars of similar metallicity, which was much smaller than their total adopted uncertainty. We expect that the systematic differences in abundances caused by using $\alpha$-enhanced models in the work presented here to be similarly negligible compared to our abundance uncertainties. Hence, for more direct comparison with literature results we have used $\alpha$-enhanced atmosphere models. Line lists were generated using the linemake package\footnote{https://github.com/vmplacco/linemake} (C. Sneden, private comm.), including molecular lines for CH, C$_2$, and CN and isotopic shift and hyperfine structure information. Representative errors arising from stellar parameter uncertainties computed for J220423 are listed in Table \ref{tab:error}. These were determined by deriving abundances for the star using different atmospheric models, each with one parameter varied by its uncertainty as given in Table \ref{tab:params}. The uncertainties were then added in quadrature including covariance terms following \citet{mcwilliam2013} and \citet{johnson2002} to provide the systematic uncertainty, $\sigma_{sys}$, on $\mathrm{[X/H]}$. The covariances were computed using the following equation
\begin{equation}
\sigma_{XY}=\frac{1}{N}\sum_i^N(X_i-\bar{X})(Y_i-\bar{Y})    
\end{equation}{}
To determine $\sigma_{T\log g}$, $\sigma_{T\xi}$, and $\sigma_{T\mathrm{[M/H]}}$, 20 model atmospheres with effective temperatures drawn from a Gaussian distribution with a mean of 4585~K and standard deviation of 150~K were computed. $\log g$ and $\xi$ was then varied in turn until we obtained ionization equilibrium between the \ion{Fe}{1} and \ion{Fe}{2} lines for $\sigma_{T\log g}$, and no trend in line abundances with reduced EW of \ion{Fe}{1} lines for $\sigma_{T\xi}$, while the direct change in $\mathrm{[Fe/H]}$ was used for $\sigma_{T\mathrm{[M/H]}}$. In the case of $\sigma_{\log g\xi}$, 20 model atmospheres with microturbulences drawn from a Gaussian distribution with a mean of 2.25~km~s$^{-1}$ and $\sigma$ of 0.3~km~s$^{-1}$ were computed and the gravity was then again varied to obtain ionization equilibrium between the \ion{Fe}{1} and \ion{Fe}{2} lines. The final covariances resulting from this process are $\sigma_{T\log g}$=49, $\sigma_{T\xi}$=13, $\sigma_{T\mathrm{[M/H]}}$=20, and $\sigma_{\log g\xi}$=-0.02.

\begin{deluxetable}{llrrrrrr}
\tabletypesize{\scriptsize}
\tablecaption{EW and atomic data for \ion{Fe}{1} and \ion{Fe}{2} lines used for parameter determination. \label{tab:ew}}
\tablewidth{0pt}
\tablehead{\colhead{Stellar ID} & \colhead{Species} & \colhead{$\lambda$} &
\colhead{$\chi$} & \colhead{$\log$ $gf$} & \colhead{EW} &
\colhead{$\sigma_{\rm EW}$} & \colhead{$\log \epsilon$}\\ 
\colhead{} & \colhead{} & \colhead{(\AA)} & \colhead{(eV)} & \colhead{} &
\colhead{(m\AA)} & \colhead{(m\AA)} & \colhead{}}
\startdata
J220409 &\ion{Fe}{1} &  3876.04 &  1.010 & $-$2.89  &  80.1 & 13.4 &  5.01\\
J220409 &\ion{Fe}{1} &  4001.66 &  2.174 & $-$1.90  &  51.4 & 13.3 &  4.91\\
J220409 &\ion{Fe}{1} &  4009.71 &  2.221 & $-$1.25  &  64.6 & 12.1 &  4.52\\
J220409 &\ion{Fe}{1} &  4084.49 &  3.329 & $-$0.54  &  55.7 & 12.8 &  4.97\\
J220409 &\ion{Fe}{1} &  4132.90 &  2.840 & $-$0.92  &  41.6 & 10.7 &  4.54\\
J220409 &\ion{Fe}{1} &  4147.67 &  1.484 & $-$2.07  &  84.6 & 13.7 &  4.78\\
J220409 &\ion{Fe}{1} &  4152.17 &  0.957 & $-$3.16  &  53.2 & 10.8 &  4.72\\
J220409 &\ion{Fe}{1} &  4153.90 &  3.397 & $-$0.28  &  37.6 &  9.2 &  4.48\\
J220409 &\ion{Fe}{1} &  4156.80 &  2.830 & $-$0.81  &  46.2 & 11.5 &  4.49\\
\multicolumn{1}{c}{\vdots} & \multicolumn{1}{l}{\phantom{0}\vdots} & \multicolumn{1}{c}{\vdots} & \multicolumn{1}{c}{\vdots} & \multicolumn{1}{c}{\vdots} & \multicolumn{1}{c}{\vdots} & \multicolumn{1}{c}{\vdots} \\
J220409 &\ion{Fe}{2} &  4508.29 &  2.860 & $-$2.25 &   48.2 & 13.9 &  4.67\\
J220409 &\ion{Fe}{2} &  4515.34 &  2.840 & $-$2.36 &   52.3 & 14.7 &  4.81\\
J220409 &\ion{Fe}{2} &  4520.22 &  2.810 & $-$2.60 &   41.6 & 13.9 &  4.85\\
J220409 &\ion{Fe}{2} &  4522.63 &  2.840 & $-$1.99 &   93.7 & 18.6 &  5.12\\
J220409 &\ion{Fe}{2} &  4541.52 &  2.860 & $-$2.79 &   38.1 & 15.4 &  5.04\\
\enddata
\tablecomments{
The complete version of Table~\ref{tab:ew} is available online only.
A short version is shown here to illustrate its form and content.
}
\end{deluxetable}

\begin{deluxetable}{llrrrrrr}
\tabletypesize{\scriptsize}
\tablecaption{EW measurements and atomic data for lines used for abundance determination. \label{tab:abunew}}
\tablewidth{0pt}
\tablehead{\colhead{Stellar ID} & \colhead{Species} & \colhead{$\lambda$} &
\colhead{$\chi$} & \colhead{$\log$ $gf$} & \colhead{EW} &
\colhead{$\sigma_{\rm EW}$} & \colhead{$\log \epsilon$}\\ 
\colhead{} & \colhead{} & \colhead{(\AA)} & \colhead{(eV)} & \colhead{} & \colhead{(m\AA)} & \colhead{(m\AA)} &\colhead{}}
\startdata
J220409 &\ion{Mg}{1} &  3986.75 &  4.343 & $-$1.44  &  21.7  &  9.7 & 5.38\\
J220409 &\ion{Mg}{1} &  4057.51 &  4.343 & $-$1.20  &  40.1  & 10.4 & 5.49\\
J220409 &\ion{Mg}{1} &  4167.27 &  4.343 & $-$1.00  &  50.2  & 12.7 & 5.44\\
J220409 &\ion{Mg}{1} &  4571.10 &  0.000 & $-$5.69  &  87.1  & 16.8 & 5.46\\
J220409 &\ion{Mg}{1} &  4702.99 &  4.343 & $-$0.67  &  75.4  & 15.0 & 5.41\\
J220409 &\ion{Ca}{1} &  4283.01 &  1.884 & $-$0.22  &  52.5  & 12.2 & 3.68\\
J220409 &\ion{Ca}{1} &  4318.65 &  1.897 & $-$0.21  &  62.0  & 13.3 & 3.83\\
J220409 &\ion{Ca}{1} &  4425.44 &  1.878 & $-$0.39  &  28.6  & 11.3 & 3.43\\
J220409 &\ion{Ca}{1} &  4455.89 &  1.897 & $-$0.51  &  27.2  & 10.9 & 3.54\\
J220409 &\ion{Ca}{1} &  5594.46 &  2.521 & $-$0.05  &  38.8  & 13.3 & 3.97\\
\enddata
\tablecomments{
The complete version of Table~\ref{tab:abunew} is available online only.
A short version is shown here to illustrate its form and content.
}
\end{deluxetable}

\section{Results \label{sec:results}}
Abundances or upper limits have been derived for 31 elements from C to Er in the three spectroscopically confirmed members of Gru~II. All abundances and upper limits are presented in Table \ref{tab:abun} listing the $\log_\epsilon$(X), $\mathrm{[X/H]}$, and $\mathrm{[X/Fe]}$ abundances along with the number of lines used to derive the abundance, standard deviation ($\sigma_{stat}$) and the total uncertainty on $\mathrm{[X/H]}$ and $\mathrm{[X/Fe]}$ calculated including covariance terms as described above. For elements were the abundances are derived from only one or two lines we use an estimated $\sigma_{\mathrm{[X/H]},stat}=0.2$ when calculating the total uncertainty.
In Figure \ref{fig:abun} we compare these abundances to results from high-resolution studies of other UFD galaxies and metal-poor MW halo stars \citep{roederer2014a}. Only abundance detections have been included in the comparison sample. The UFD galaxies are: Bo\"{o}tes~I \citep{feltzing2009,frebel2016,gilmore2013,ishigaki2014,norris2010}, Bo\"{o}tes~II \citep{ji2016b}, Coma Berenices \citep{frebel2010}, Grus~I \citep{ji2019a}, Hercules \citep{koch2008}, Horologium~I \citep{nagasawa2018}, Leo~IV \citep{simon2010}, Pisces~II \citep{spite2018}, Reticulum~II \citep{ji2016c}, Segue~1 \citep{frebel2014,norris2010}, Segue~2 \citep{roederer2014b}, Triangulum~II \citep{ji2019a}, Tucana~II \citep{ji2016d,chiti2018}, Tucana~III \citep{hansen2017,marshall2019}, and Ursa Major~II \citep{frebel2010}.

\subsection{Carbon and Odd Z Elements}
\begin{figure*}[hbt!]
\includegraphics[scale=0.35]{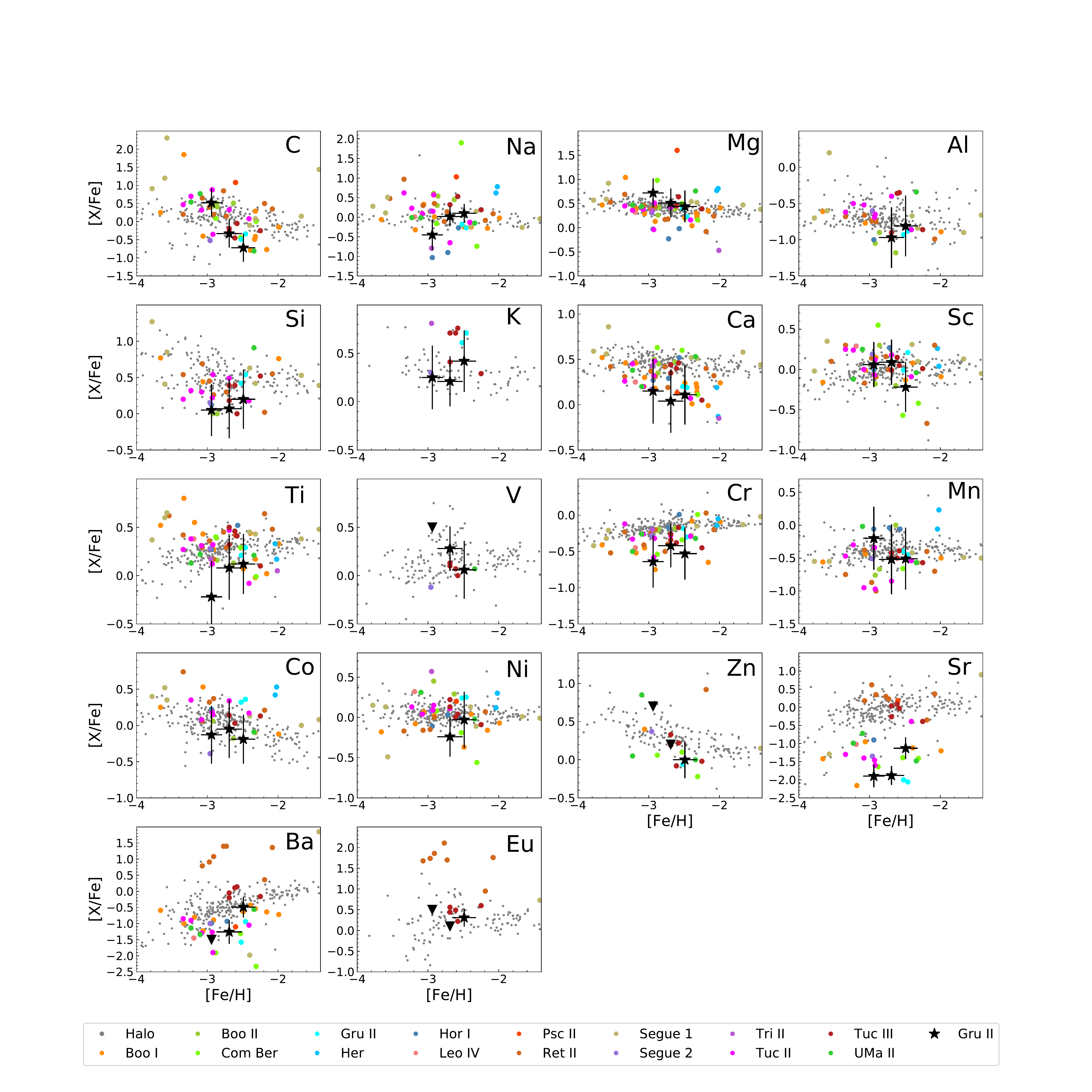}
\caption{$\mathrm{[X/Fe]}$ derived abundances for Gru~II (black stars) compared to stellar abundances from the MW halo \citep[grey dots;][]{roederer2014a} and other UFD galaxies (colored dots according to legend, see text for references). Upper limits for Gru~II stars are marked with downward pointing black triangles\label{fig:abun}.}
\end{figure*}

Carbon abundances were determined from synthesis of the CH $G$-band at 4313\,{\AA} assuming a standard oxygen enhancement for metal-poor stars of $\mathrm{[O/Fe]}=0.4$ as a direct oxygen measurement could not be made. Na abundances were derived from the 5889{\AA} and 5895{\AA} lines via synthesis in J220409 and J220423, and EW analysis in J220352. Abundances for Al could only be determined for J220409 and J220423 and were determined via synthesis of the 3944{\AA} and 3961{\AA} lines. The 7664{\AA} and 7698{\AA} lines were used to derive the K abundances using EWs in J220423 and synthesis in J220409 and J220352. Finally a mixture of the Sc lines at 4226{\AA}, 4320{\AA}, 4400{\AA}, 4415{\AA}, 4670{\AA}, 5526{\AA}, and 5658{\AA} was synthesized to derive Sc abundances in the stars. 
As can be seen in Figure \ref{fig:abun} some spread is detected in the C abundances for the three stars, however, effects of stellar evolution are known to alter the surface C abundance of low mass stars. Following \citet{placco2014} we find carbon corrections of $\Delta$C =  $+0.70$ for J220423, $+0.46$ for J220409, and $+0.07$ for J220352, reducing the spread in the birth carbon abundance of the stars. Taking these corrections into account, none of the stars qualify as Carbon Enhanced Metal-Poor (CEMP) stars ($\mathrm{[C/Fe]} > 0.7$; \citet{aoki2007}). In Figure \ref{fig:abun} we plot the original C abundances for easy reference to the literature values. It should be noted that many of the stars in the Figure are giants thus their original C abundance may be somewhat different from the measured C abundances plotted. For Na, Al, and Sc we find abundances of the three Gru~II stars that are similar to other UFD and metal-poor halo stars (see Figure \ref{fig:abun}). K abundances have only been derived for stars in a few UFD galaxies but the majority of these seem to cluster around $\mathrm{[K/Fe]}=0.7$ while the stars in Gru~II exhibit lower K abundances around $\mathrm{[K/Fe]}=0.3$, closer to the abundances found in the halo comparison sample \citep{roederer2014a}. 

\begin{deluxetable*}{lrrrrrrrrrrrrrrrrrrrrr}
\rotate
\tabletypesize{\scriptsize}
\tablecaption{Abundances and associated uncertainties derived for the three stars.\label{tab:abun}} 
\tablehead{
\multicolumn{1}{c}{}&\multicolumn{7}{c}{J220409}&\multicolumn{7}{c}{J220423}&\multicolumn{7}{c}{J220351}\\
\colhead{Element} & \colhead{$\log\epsilon$(X)} &  \colhead{N} &
\colhead{$\sigma_{\rm stat}$} & \colhead{$\mathrm{[X/H]}$} &
\colhead{$\sigma_{\rm [X/H]}$} & \colhead{$\mathrm{[X/Fe]}$} &
\colhead{$\sigma_{\rm [X/Fe]}$} & \colhead{$\log\epsilon$(X)} &  \colhead{N} &
\colhead{$\sigma_{\rm stat}$} & \colhead{$\mathrm{[X/H]}$} &
\colhead{$\sigma_{\rm [X/H]}$} & \colhead{$\mathrm{[X/Fe]}$} &
\colhead{$\sigma_{\rm [X/Fe]}$} & \colhead{$\log\epsilon$(X)} &  \colhead{N} &
\colhead{$\sigma_{\rm stat}$} & \colhead{$\mathrm{[X/H]}$} &
\colhead{$\sigma_{\rm [X/H]}$} & \colhead{$\mathrm{[X/Fe]}$} &
\colhead{$\sigma_{\rm [X/Fe]}$} \\
\colhead{} & \colhead{(dex)} &\colhead{} & \colhead{(dex)} & \colhead{(dex)}
&\colhead{(dex)} & \colhead{(dex)} &\colhead{(dex)} & \colhead{(dex)} &\colhead{} & \colhead{(dex)} & \colhead{(dex)}
&\colhead{(dex)} & \colhead{(dex)} &\colhead{(dex)} & \colhead{(dex)}
&\colhead{} & \colhead{(dex)} & \colhead{(dex)} &\colhead{(dex)} & \colhead{(dex)} &\colhead{(dex)}}
\startdata
C    &   5.41 &  &   0.20 & $-$3.02&0.47    & $-$0.33 &0.31    &    5.22&   &   0.20 & $-$3.21&0.47    & $-$0.72 &0.32    &    6.01&  &    0.20& $-$2.44&0.47    &    0.52 &0.30 \\   
N    &$<+$7.83&  &$\cdots$&$<-$0.69&$\cdots$&$<+$2.00 &$\cdots$&$<+$7.34&   &$\cdots$&$<-$0.49&$\cdots$&$<+$2.00 &$\cdots$&$<+$6.89&  &$\cdots$&$<-$0.94&$\cdots$&$<+$2.00 &$\cdots$ \\
NaI  &   3.57 &2 &   0.20 & $-$2.67&0.30    &    0.02 &0.27    &    3.85& 2 &   0.20 & $-$2.39&0.30    &    0.10 &0.28    &    2.85&2 &    0.32& $-$3.39&0.39    & $-$0.45 &0.36 \\ 
MgI  &   5.44 &5 &   0.17 & $-$2.18&0.27    &    0.51 &0.25    &    5.55& 5 &   0.21 & $-$2.05&0.30    &    0.44 &0.29    &    5.38&2 &    0.20& $-$2.22&0.29    &    0.72 &0.26 \\    
AlI  &   2.79 &2 &   0.20 & $-$3.66&0.50    & $-$0.97 &0.32    &    3.15& 2 &   0.20 & $-$3.30&0.46    & $-$0.81 &0.26    &$\cdots$&  &$\cdots$&$\cdots$&$\cdots$&$\cdots$ &$\cdots$ \\
SiI  &   4.89 &1 &   0.20 & $-$2.62&0.43    &    0.07 &0.28    &    5.22& 1 &   0.20 & $-$2.29&0.43    &    0.20 &0.38    &    4.62&2 &    0.20& $-$2.89&0.43    &    0.05 &0.26 \\    
KI   &   2.55 &2 &   0.20 & $-$2.48&0.30    &    0.21 &0.27    &    2.96& 2 &   0.20 & $-$2.07&0.30    &    0.42 &0.28    &    2.34&1 &    0.20& $-$2.69&0.30    &    0.25 &0.26 \\    
CaI  &   3.69 &13&   0.26 & $-$2.65&0.32    &    0.04 &0.33    &    3.96& 13&   0.24 & $-$2.38&0.30    &    0.11 &0.32    &    3.55&5 &    0.27& $-$2.79&0.33    &    0.15 &0.33 \\    
ScII &   0.55 &6 &   0.09 & $-$2.60&0.24    &    0.09 &0.23    &    0.44& 5 &   0.16 & $-$2.71&0.27    & $-$0.22 &0.28    &    0.27&2 &    0.20& $-$2.88&0.30    &    0.06 &0.28 \\    
TiI  &   2.33 &4 &   0.23 & $-$2.62&0.38    &    0.07 &0.29    &    2.48& 8 &   0.20 & $-$2.47&0.36    &    0.02 &0.27    &$\cdots$&  &$\cdots$&$\cdots$&$\cdots$&$\cdots$ &$\cdots$ \\
TiII &   2.34 &11&   0.25 & $-$2.61&0.29    &    0.08 &0.33    &    2.58& 19&   0.22 & $-$2.37&0.26    &    0.12 &0.32    &    1.79&7 &    0.29& $-$3.16&0.32    & $-$0.22 &0.36 \\    
VII  &   1.52 &2 &   0.20 & $-$2.41&0.26    &    0.28 &0.30    &    1.50& 1 &   0.20 & $-$2.43&0.26    &    0.06 &0.31    &$<+$1.73&  &$\cdots$&$<-$2.44&$\cdots$&$<+$0.50 &$\cdots$ \\ 
CrI  &   2.53 &4 &   0.10 & $-$3.11&0.27    & $-$0.42 &0.21    &    2.62& 7 &   0.22 & $-$3.02&0.34    & $-$0.53 &0.29    &    2.06&2 &    0.23& $-$3.58&0.34    & $-$0.64 &0.28 \\    
MnI  &   2.22 &3 &   0.30 & $-$3.21&0.56    & $-$0.52 &0.39    &    2.43& 7 &   0.17 & $-$3.00&0.50    & $-$0.51 &0.31    &    2.29&1 &    0.20& $-$3.14&0.52    & $-$0.20 &0.31 \\    
FeI  &   4.82 &94&   0.17 & $-$2.69&0.34    &$\cdots$ &$\cdots$&    5.01&116&   0.18 & $-$2.49&0.34    & $\cdots$&$\cdots$&    4.57&32&    0.15& $-$2.94&0.33    &$\cdots$ &$\cdots$ \\ 
FeII &   4.81 &10&   0.21 & $-$2.76&0.25    &$\cdots$ &$\cdots$&    5.01&14 &   0.16 & $-$2.49&0.20    & $\cdots$&$\cdots$&    4.56&10&    0.25& $-$2.95&0.28    &$\cdots$ &$\cdots$ \\
CoI  &   2.25 &1 &   0.20 & $-$2.74&0.39    & $-$0.05 &0.27    &    2.31& 2 &   0.20 & $-$2.68&0.39    & $-$0.19 &0.28    &    1.92&1 &    0.20& $-$3.07&0.39    & $-$0.13 &0.26 \\    
NiI  &   3.29 &2 &   0.20 & $-$2.93&0.21    & $-$0.24 &0.29    &    3.70& 7 &   0.26 & $-$2.52&0.32    & $-$0.03 &0.34    &$\cdots$&  &$\cdots$&$\cdots$&$\cdots$&$\cdots$ &$\cdots$ \\
CuI  &$<+$1.40&  &$\cdots$&$<-$2.79&$\cdots$&$<-$0.10 &$\cdots$&$<+$1.40&   &$\cdots$&$<-$2.79&$\cdots$&$<-$0.30 &$\cdots$&$<+$2.05&  &$\cdots$&$<-$2.14&$\cdots$&$<+$0.80 &$\cdots$ \\ 
ZnI  &$<+$2.07&  &$\cdots$&$<-$2.49&$\cdots$&$<+$0.20 &$\cdots$&    2.07& 2 &   0.20 & $-$2.49&0.29    &    0.00 &0.30    &$<+$2.32&  &$\cdots$&$<-$2.24&$\cdots$&$<+$0.70 &$\cdots$ \\ 
SrII & $-$1.70&2 &   0.20 & $-$4.57&0.29    & $-$1.88 &0.31    & $-$0.75& 2 &   0.20 & $-$3.62&0.29    & $-$1.13 &0.32    & $-$1.97&1 &    0.20& $-$4.84&0.29    & $-$1.90 &0.31 \\    
YII  &$<-$0.78&  &$\cdots$&$<-$2.99&$\cdots$&$<-$0.30 &$\cdots$& $-$1.18& 1 &   0.20 & $-$3.39&0.23    & $-$0.90 &0.33    &$<-$0.73&  &$\cdots$&$<-$2.94&$\cdots$&$<+$0.00 &$\cdots$ \\  
ZrII &$<-$0.61&  &$\cdots$&$<-$3.19&$\cdots$&$<-$0.50 &$\cdots$& $-$0.41& 1 &   0.20 & $-$2.99&0.30    & $-$0.50 &0.29    &$<-$0.36&  &$\cdots$&$<-$2.94&$\cdots$&$<+$0.00 &$\cdots$ \\  
BaII & $-$1.77&3 &   0.19 & $-$3.95&0.38    & $-$1.26 &0.26    & $-$0.80& 3 &   0.06 & $-$2.98&0.33    & $-$0.49 &0.19    &$<-$2.26&  &$\cdots$&$<-$4.44&$\cdots$&$<-$1.50 &$\cdots$ \\  
LaII &$<-$1.59&  &$\cdots$&$<-$2.69&$\cdots$&$<+$0.00 &$\cdots$& $-$1.21& 2 &   0.20 & $-$2.31&0.53    &    0.18 &0.33    &$<-$1.04&  &$\cdots$&$<-$2.14&$\cdots$&$<+$0.80 &$\cdots$ \\  
CeII &$<-$0.91&  &$\cdots$&$<-$2.49&$\cdots$&$<+$0.20 &$\cdots$&$<-$0.61&   &$\cdots$&$<-$2.19&$\cdots$&$<+$0.30 &$\cdots$&$<-$0.56&  &$\cdots$&$<-$2.14&$\cdots$&$<+$0.80 &$\cdots$ \\  
PrII &$<-$1.17&  &$\cdots$&$<-$1.89&$\cdots$&$<+$0.80 &$\cdots$& $-$1.48& 3 &   0.03 & $-$2.20&0.26    &    0.29 &0.20    &$\cdots$&  &$\cdots$&$\cdots$&$\cdots$&$\cdots$ &$\cdots$ \\  
NdII &$<-$0.57&  &$\cdots$&$<-$1.99&$\cdots$&$<+$0.70 &$\cdots$& $-$0.99& 4 &   0.07 & $-$2.41&0.24    &    0.08 &0.23    &$<-$0.58&  &$\cdots$&$<-$2.04&$\cdots$&$<+$0.90 &$\cdots$ \\  
SmII &$\cdots$&  &$\cdots$&$\cdots$&$\cdots$&$\cdots$ &$\cdots$&$<-$1.03&   &$\cdots$&$<-$1.99&$\cdots$&$<+$0.50 &$\cdots$&$<-$0.88&  &$\cdots$&$<-$1.84&$\cdots$&$<+$1.10 &$\cdots$ \\  
EuII &$<-$2.07&  &$\cdots$&$<-$2.59&$\cdots$&$<+$0.10 &$\cdots$& $-$1.66& 3 &   0.10 & $-$2.18&0.21    &    0.31 &0.25    &$<-$1.92&  &$\cdots$&$<-$1.44&$\cdots$&$<+$0.50 &$\cdots$ \\  
GdII &$<-$0.92&  &$\cdots$&$<-$1.99&$\cdots$&$<+$0.70 &$\cdots$&$<-$0.62&   &$\cdots$&$<-$1.69&$\cdots$&$<+$0.80 &$\cdots$&$\cdots$&  &$\cdots$&$\cdots$&$\cdots$&$\cdots$ &$\cdots$ \\  
DyII &$<-$0.59&  &$\cdots$&$<-$1.69&$\cdots$&$<+$1.00 &$\cdots$&$<-$0.69&   &$\cdots$&$<-$1.79&$\cdots$&$<+$0.70 &$\cdots$&$\cdots$&  &$\cdots$&$\cdots$&$\cdots$&$\cdots$ &$\cdots$ \\  
ErII &$<-$1.27&  &$\cdots$&$<-$1.19&$\cdots$&$<+$0.50 &$\cdots$&$<-$1.07&   &$\cdots$&$<-$1.99&$\cdots$&$<+$0.50 &$\cdots$&$\cdots$&  &$\cdots$&$\cdots$&$\cdots$&$\cdots$ &$\cdots$ \\  
\enddata
\end{deluxetable*} 

\begin{deluxetable}{lccccc}
\tablecaption{Abundance errors arising from stellar parameter
uncertainties for DES J220423.\label{tab:error}} 
\tablehead{\colhead{Element}   & \colhead{$\Delta$T$_{eff}$}   &   \colhead{$\Delta\log g$}   &   \colhead{$\Delta\xi$}   &   \colhead{$\Delta\mathrm{[M/H]}$}   &   \colhead{$\sigma_{sys}$}\\
\colhead{} &  \colhead{(dex)}   & \colhead{(dex)}   &   \colhead{(dex)}   &  \colhead{(dex)}   &   \colhead{(dex)} }
\startdata
CH              & 0.27 &   0.12 &   0.04 & 0.04 & 0.43   \\
\ion{Na}{1}     & 0.15 &   0.03 &   0.10 & 0.00 & 0.22   \\
\ion{Mg}{1}     & 0.14 &   0.04 &   0.07 & 0.01 & 0.21   \\
\ion{Al}{1}     & 0.22 &   0.13 &   0.07 & 0.14 & 0.45   \\
\ion{Si}{1}     & 0.25 &   0.05 &   0.15 & 0.02 & 0.38   \\
\ion{K}{1}      & 0.17 &   0.02 &   0.06 & 0.01 & 0.22   \\
\ion{Ca}{1}     & 0.13 &   0.03 &   0.05 & 0.01 & 0.18   \\
\ion{Sc}{1}     & 0.08 &   0.04 &   0.17 & 0.00 & 0.22   \\
\ion{Ti}{1}     & 0.25 &   0.03 &   0.05 & 0.01 & 0.30   \\
\ion{Ti}{2}     & 0.04 &   0.09 &   0.06 & 0.01 & 0.14   \\
\ion{V}{2}      & 0.05 &   0.10 &   0.02 & 0.02 & 0.16   \\
\ion{Cr}{1}     & 0.22 &   0.02 &   0.02 & 0.01 & 0.25   \\
\ion{Mn}{1}     & 0.30 &   0.05 &   0.10 & 0.12 & 0.47   \\
\ion{Fe}{1}     & 0.21 &   0.03 &   0.10 & 0.02 & 0.29   \\
\ion{Fe}{2}     & 0.02 &   0.10 &   0.05 & 0.01 & 0.13   \\
\ion{Co}{1}     & 0.25 &   0.03 &   0.07 & 0.04 & 0.33   \\
\ion{Ni}{1}     & 0.16 &   0.01 &   0.02 & 0.01 & 0.18   \\
\ion{Zn}{1}     & 0.10 &   0.10 &   0.00 & 0.00 & 0.21   \\
\ion{Sr}{2}     & 0.02 &   0.02 &   0.20 & 0.05 & 0.21   \\
\ion{Y}{2}      & 0.07 &   0.03 &   0.03 & 0.00 & 0.11   \\
\ion{Zr}{2}     & 0.09 &   0.13 &   0.04 & 0.00 & 0.23   \\
\ion{Ba}{2}     & 0.13 &   0.16 &   0.08 & 0.05 & 0.33   \\
\ion{La}{2}     & 0.20 &   0.22 &   0.11 & 0.10 & 0.49   \\
\ion{Pr}{2}     & 0.16 &   0.07 &   0.03 & 0.04 & 0.26   \\
\ion{Nd}{2}     & 0.12 &   0.08 &   0.00 & 0.04 & 0.22   \\
\ion{Eu}{2}     & 0.09 &   0.08 &   0.01 & 0.02 & 0.18   \\
\enddata
\end{deluxetable}

\subsection{$\alpha$ and Iron-Peak Elements}
EW analysis was used to derived Mg and Ti abundances in all three stars. Ca abundances were also derived from EW analysis in J220409 and J220423 while spectral synthesis of the Ca lines at 4426{\AA}, 4434{\AA}, 4454{\AA}, 6122{\AA}, and 6162{\AA} was used to determine the Ca abundance in J220352. Cr and Ni abundances were determined using EW analysis in J220423, while synthesis of the Cr lines at 4252{\AA}, 4274{\AA}, 4289{\AA}, 4646{\AA}, and 5206{\AA} and Ni lines at 3807{\AA}, 5476{\AA}, and 6644{\AA} was used in J220409 and J220352. For the remaining elements in this group, Si, V, Mn, Co, Cu, and Zn, abundances or upper limits were determined via spectral synthesis in all three stars using the following lines: Si, 3905{\AA} and 4102{\AA}; V, 3952{\AA} and 4005{\AA}; Mn, 4030{\AA}, 4033{\AA}, 4034{\AA}, 4041{\AA}, 4754{\AA}, 4823{\AA}, and 4762{\AA}; Co, 4118{\AA} and 4121{\AA}; Cu, 5105{\AA}; and Zn, 4722{\AA} and 4810{\AA}.
All three stars are enhanced in Mg, similar to other metal-poor stars in the halo and UFDs. However the derived abundances for Si, Ca, and Ti, which are also usually enhanced in metal-poor stars, are low in all three Gru~II stars compared to the halo and UFD galaxy sample. Mg, Si, Ca, and Ti are primarily created in massive (M $>8$M$_\odot$) stars, with Si, Ca, and Ti being synthesized during the explosive nucleosynthesis in the CCSN phase and Mg being created during the hydrostatic burning phases of the stars \citep{woosley1995}. A high ratio of hydrostatic to explosive $\alpha$-elements as observed in these stars can be produced in high mass, $\geqslant20$M$_\odot$ CCSN \citep{heger2010}. The three Gru~II stars also exhibit low Cr abundances, a feature in common with stars in other UFD galaxies and also found for the most metal-poor MW halo stars \citep{mcwilliam1995}. The abundances derived for V, Mn, Co, Ni, and Zn in the Gru~II stars follow the abundance trends observed in other UFDs and metal-poor halo stars (see Figure \ref{fig:abun}).

\subsection{Neutron-Capture Elements}
Abundances for all neutron-capture elements were derived from spectral synthesis. It was only possible to derive abundances for Sr in all three stars using the 4077{\AA} and 4215{\AA} lines. Ba abundances for J220904 and J220423 and the upper limit in J220352 were derived from the 5853{\AA}, 6141{\AA}, and 6496{\AA} lines. The most metal-rich of the three stars J220423 exhibits a small enhancement in some of the heavy neutron-capture elements ($\mathrm{[Eu/Fe]}=0.31\pm0.22$), enabling the derivation of abundances for Y (4398{\AA}), Zr (4149{\AA}), La (3995{\AA} and 4086{\AA}), Pr (4179{\AA}, 4222{\AA}, and 4408{\AA}), Nd (4109{\AA}, 4177{\AA}, 4462{\AA}, and 4825{\AA}), and Eu (3907{\AA}, 4129{\AA}, and 4205{\AA}) for this star. Eu upper limits derived from the 4129{\AA} line are given for J220904 and J220352 also. The Sr and Ba abundances derived for J220423 are sub-solar but higher than what is found in the other two stars analysed, suggesting that an additional source of neutron-capture elements has enriched this star. J220423 has $\mathrm{[Ba/Eu]}= -0.80$, compatible with an $r$-process origin of the neutron-capture element excess \citep{sneden2008}. The two non-enhanced stars J220409 and J220352 both exhibit extremely low Sr (and Ba) abundances similar to neutron-capture element abundances detected in other UFDs \citep{ji2019a}, supporting the dwarf galaxy classification of Gru~II (see Figure \ref{fig:abun}).

\section{Discussion \label{sec:discus}}
\begin{figure}[hbt!]
\centering 
\includegraphics[scale=0.5]{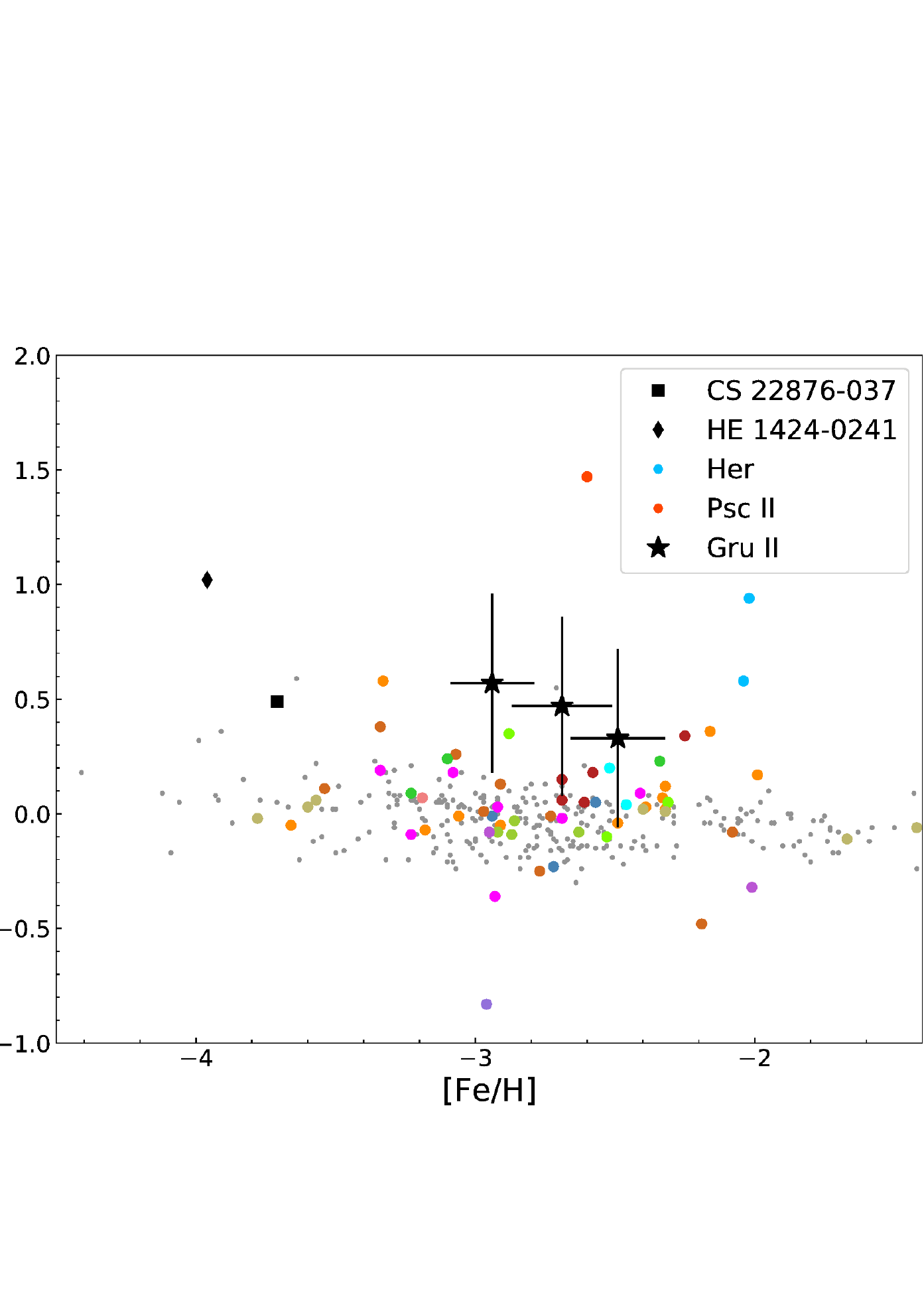}
\caption{$\mathrm{[Mg/Ca]}$ as a function of metallicity for the three Gru~II stars compared to stars in other UFDs and metal-poor MW halo stars, color as in Figure \ref{fig:abun}. The three stars in Gru~II show markedly higher $\mathrm{[Mg/Ca]}$ ratios than the majority of the comparison stars. The two halo stars with similar abundance signatures, CS~22876-037 and HE~1424$-$0241 are marked with a black square and diamond respectively. The star with the highest $\mathrm{[Mg/Ca]}$ value (orange-red dot) is a CEMP-no star in Pis~II \citep{spite2018}. 
\label{fig:MgCa}}
\end{figure}

\begin{figure}
\includegraphics[scale=0.6]{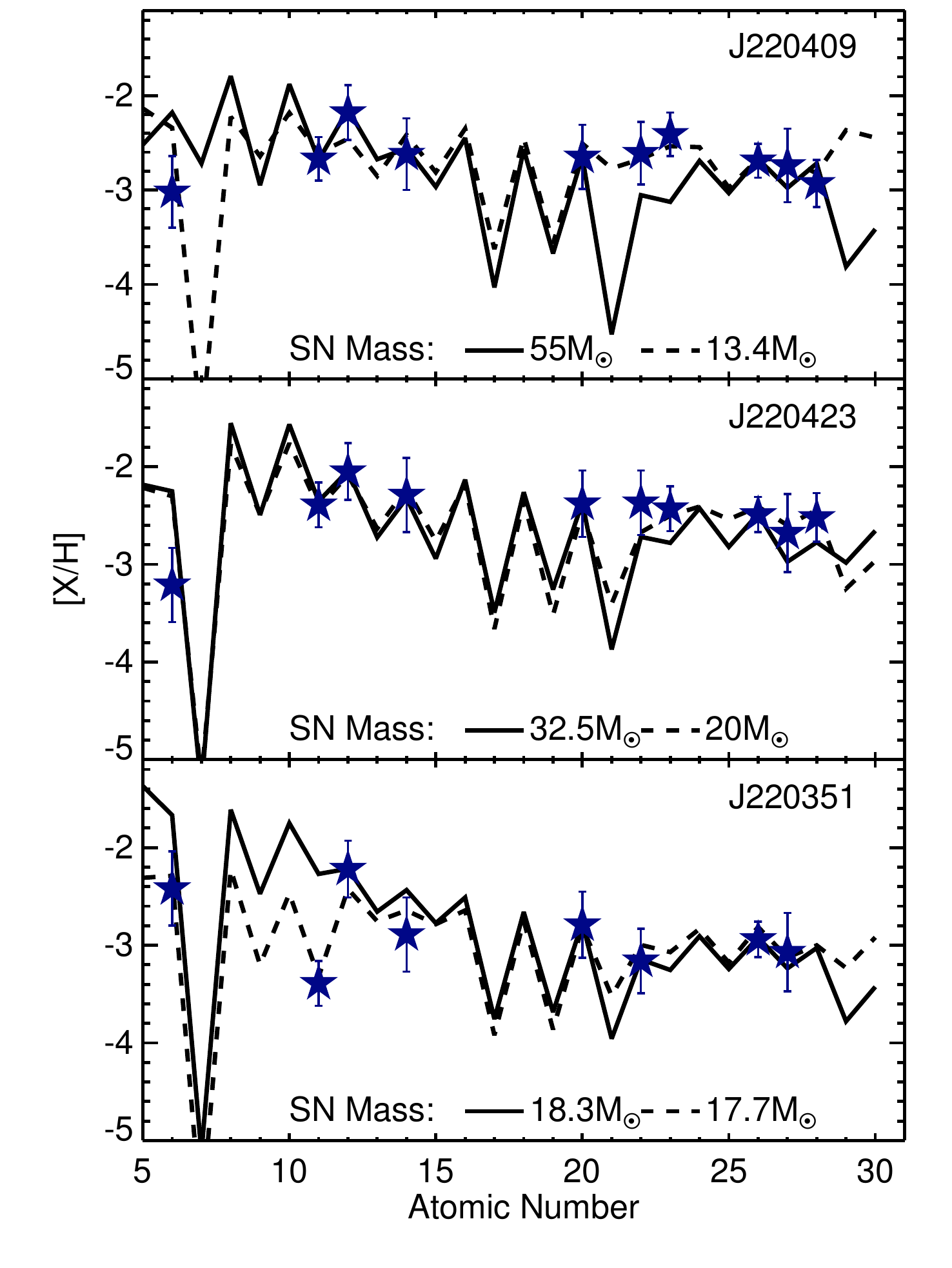}
\caption{Yields from best fit SNe models from \citet{heger2010} to the $\mathrm{[Mg/H]}$, $\mathrm{[Ca/H]}$, $\mathrm{[Fe/H]}$ abundances of the three stars analysed. Blue stars represent the derived abundances for the three stars. Solid lines show the model fits using only the $\mathrm{[Mg/H]}$, $\mathrm{[Ca/H]}$, $\mathrm{[Fe/H]}$ abundances and dashed lines shows model fits using all abundances plotted. \label{fig:SNfit}}
\end{figure}

For the majority of the elements, the abundances derived for Gru~II follow the trends detected in other UFD galaxies and in the MW halo, with a few notable outliers. In Figure \ref{fig:MgCa} we plot the $\mathrm{[Mg/Ca]}$ ratio as a function of metallicity for the three Gru~II stars along with the UFD galaxies and halo stars from Figure \ref{fig:abun}. The three stars in Gru~II clearly stand out, displaying somewhat higher $\mathrm{[Mg/Ca]}$ ratios than the majority of the comparison sample stars. This $\alpha$-element signature was first discovered in the MW halo star CS~22876-037 (black square in Figure \ref{fig:MgCa}) for which \citet{norris2000} reported abundances of $\mathrm{[Mg/Fe] = 0.5}\pm0.12$ and $\mathrm{[Ca/Fe]} = 0.01\pm0.13$. A more extreme version of this abundance signature was later observed in another MW halo star, namely HE~1424-0241 (black diamond in Figure \ref{fig:MgCa}), where \citet{cohen2007} reported abundances of $\mathrm{[Mg/Fe]} = 0.44\pm0.12$ and $\mathrm{[Ca/Fe]} = -0.58$\footnote{No uncertainty was reported on the Ca abundance as it was determined from just one absorption feature.}. Both stars display a high dominance of hydrostatic versus explosive $\alpha$-elements. From the CCSNe models of \citet{woosley1995}, \citet{norris2000} conclude that CS~22876-037 was likely enriched by a zero metallicity 30M$_\odot$ CCSN. Using the online fitting code STARFIT \footnote{\url{http://starfit.org/}} which performs a $\chi^2$ fit of SNe models to abundances data \citep[see][and subsequent online update in 2012 for more details]{heger2010}, we fit SN models to the C, Na, Mg, Ca, Ti, Fe, Co, and Ni abundances of the three Gru~II stars. Carbon values corrected for stellar evolution were used for the fit. Abundances of Sc and Cr have been excluded as these elements are generally underproduced in these models \citep{heger2010}. The result is shown in Figure \ref{fig:SNfit} were dashed lines represents the best fit using all elements and solid lines represents fits using only the Mg, Ca, and Fe abundances, demonstrating that higher mass models are preferred when only considering the Mg, Ca and Fe abundances detected in these stars. This agrees with the results from \citet{norris2000} which suggest that stars with this chemical signature were likely enriched by a population of very high-mass stars. It is notable that in Figure \ref{fig:MgCa}, stars in two other systems also stand out, namely Her (blue points) and Psc~II (orange-red point). The high $\mathrm{[Mg/Ca]}$ in the star in Psc~II is driven by a very high Mg abundance (see Figure \ref{fig:abun}). Furthermore, this star also exhibits a high C abundance and is classified as a CEMP-no star \citep{spite2018}. The source of carbon in CEMP-no stars is still debated, but rotation and mixing in the progenitor stars is likely to play a significant role.  This has also been shown to influence the production of other light elements such as Mg \citep{maeder2015}. Hence it is not clear if $\mathrm{[Mg/Ca]}$ ratios in CEMP-no and non-carbon enhanced stars, like the Gru~II stars, should be discussed in the same context. In Her, an $\alpha$-element abundance signature similar to that detected in Gru~II is seen in both stars analysed by \citet{koch2008}, who found $\mathrm{[Mg/Ca]}=0.94$ and $0.58$ for the two stars. Hence, like the stars in Gru~II, the stars in Her were also likely enriched by a population of high-mass stars. \citet{koch2008} suggest that this is either the result of an IMF of Her skewed towards higher mass stars or stochastic chemical evolution in this galaxy. In the following, we discuss these two and alternative scenarios to explain the abundance signature of Gru~II.

\subsection{Top Heavy Initial Mass Function}
Some correlations have been detected between IMFs and galaxy properties, with the largest galaxies having bottom-heavy IMFs and smaller galaxies having more bottom-light IMFs \citep[][although \citet{gennaro2018} find a more complicated picture]{geha2013}. As previously described, UFD are ideal laboratories for studying the low mass stellar IMF. \citet{geha2013} determined the IMF for Her based on Hubble Space Telescope imaging and detected a slope of $\alpha=1.2$. This is a somewhat shallower slope than the Salpeter \citep[$\alpha = 2.35$;][]{salpeter1955} or Kroupa \citep[$\alpha = 2.3$;][]{kroupa2001} IMF detected for more massive systems. On the other hand, this result is in good agreement with the high $\mathrm{[Mg/Ca]}$ abundance signature of the galaxy, and hence an overall chemical enrichment dominated by high mass stars. Again, it should be noted that the result of \citet{geha2013} is based on the low mass (M$<0.8$M$_\odot$) stellar population observable in Her today and does not necessarily directly translate to the IMF of higher mass stellar populations previously present in the galaxy. With only three stars analysed in Gru~II it is difficult to conclude if the overall chemical enrichment of the galaxy is dominated by high-mass stars. However, the very consistent $\alpha$-element signature detected in these three stars, requiring nucleosynthesis in $\geqslant$20M$_\odot$ stars to produce, is compatible with a top-heavy IMF for the Population III stars, leading to a different chemical signature in this galaxy than that seen in other UFD galaxies. It should be noted that \citet{geha2013} also measured the IMF for Leo IV and found a slope of $\alpha = 1.3$ similar to Her. Chemical abundances from high resolution data have only been presented for one star in Leo IV \citep{simon2010}. This star shows roughly equal, small Mg and Ca enhancements with $\mathrm{[Mg/Ca]}=0.07$, thus not similar to the $\alpha$-elements signature found in Her and Gru~II.

\subsection{Stochastic Chemical Enrichment}
Another possible explanation for the chemically peculiar stars in Gru~II is that the chemical enrichment and star formation in systems like Gru~II are stochastic and inhomogeneous. It can be seen in Figure \ref{fig:MgCa} that other UFD galaxy stars exhibit  a similar $\alpha$-element signature as the three Gru~II and two Her stars, including the $r$-process enhanced galaxies Ret~II and Tuc~III \citep{ji2016c,marshall2019}. In these systems, however, the high $\mathrm{[Mg/Ca]}$ stars are outliers and the majority of the stars exhibit enhancements in the $\alpha$ elements similar to metal-poor halo stars. Since the study of \citet{koch2008}, \citet{francois2016} have derived Mg and Ca abundances for an additional four stars in Her from medium resolution spectra (data not included in Figure \ref{fig:abun} and \ref{fig:MgCa}). The stars in this sample are more metal-poor ($-2.83 < \mathrm{[Fe/H]} < -2.28$) and only one of the four stars show an $\alpha$-elements signature  ($\mathrm{[Mg/Ca]} = 0.24$) similar to the \citet{koch2008} stars. Including these stars, the chemical signature of Her more resembles that of the other UFD galaxies, suggesting that stochastic and inhomogeneous chemical enrichment of Her is more likely the explanation for the varying abundance patterns detected in Her. For the abundance signature to be the result of stochastic chemical enrichment of a given galaxy, only a few chemical enrichment events can have polluted the galaxy overall. The more enrichment events a system encounters, the more washed-out the signature of any individual event becomes. \citet{koch2008} performed a stochastic sampling following \citet{mcwilliam1999} and found that the abundance ratios of their two Her stars could only reasonably be obtained in a system with fewer than 11 SNe. They note that this is somewhat at odds with the iron abundances of their stars of $\mathrm{[Fe/H]} \sim -2$, which likely requires on the order of 100 SNe to obtain \citep{koch2008}. The halo stars exhibiting the $\alpha$-element signature are found at very low metallicity ($\mathrm{[Fe/H]} < -3.5$), making it more likely that the gas from which these formed was enriched by just one SN. The three Gru~II stars have metallicities between $-2.94 < \mathrm{[Fe/H]} < -2.49$, thus it is not unlikely that the gas from which these stars formed was enriched by a low number of supernovae. Further determination of $\alpha$-abundances in a larger sample of Gru~II stars covering a larger range in metallicity would help to map out the chemical enrichment of this galaxy.

\subsection{Other Alternatives}
An alternative explanation for the low $\alpha$ abundances detected in the Gru~II stars could be an early onset of type Ia SNe. The general chemical evolution scenario predicts an enhancement in $\alpha$ elements at low metallicity as a result of enrichment by CCSNe followed by a downturn or `knee' in $\mathrm{[\alpha/Fe]}$ at the onset of type Ia SNe \citep{tinsley1979}. An early enrichment by type Ia SNe was speculated to be the reason for the small $\alpha$-element abundances detected in Hor~I \citep{nagasawa2018}. However, the three stars analysed here show a `normal' enhancement in Mg and only low Si, Ca, and Ti abundances. An injection of Fe into the systems would lower all the $\mathrm{[\alpha/Fe]}$ abundance ratios, thus we do not consider this a likely explanation for the $\alpha$-element abundance pattern detected in Gru~II.
Some UFD galaxies, including Hor~I, have also been found to be likely satellites of the Large Magellanic Clouds (LMC) rather than MW satellites \citep{kallivayalil2018}. It is not yet clear if a different birth environment like the LMC could result in the different chemical abundance patterns seen in some of the LMC associated UFD galaxies \citep{nagasawa2018}. However, as a counter-argument, \citet{simon2020} find it unlikely that Gru~II is associated with the LMC.

\subsection{Source of Neutron-Capture Elements}
\begin{figure*}[hbt!]
\centering 
\includegraphics[scale=0.4]{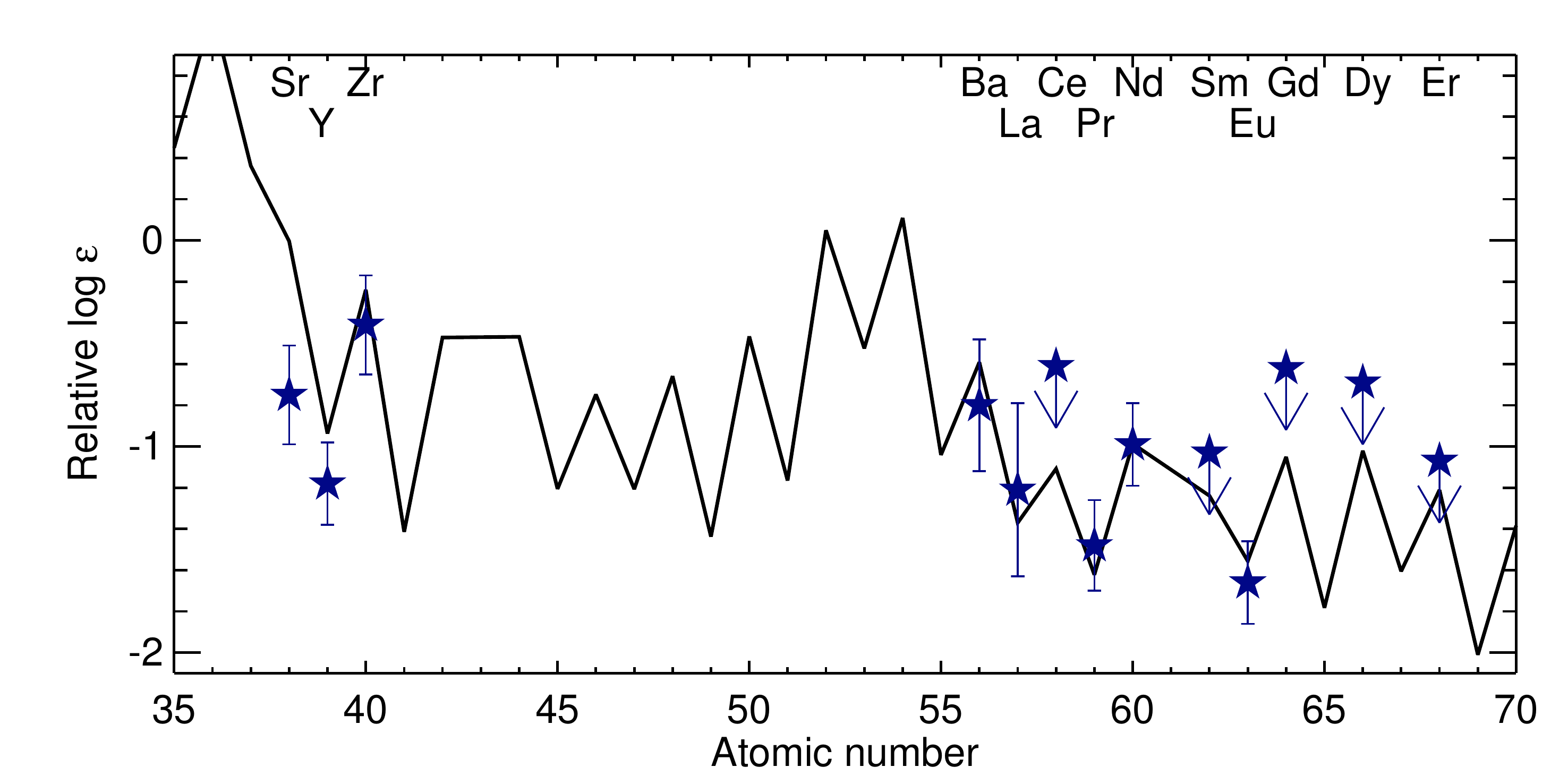}
\caption{Absolute abundances of neutron-capture elements derived for J220423 compared to the scaled solar system $r$-process abundance pattern from \citet{arlandini1999}. \label{fig:rproc}
}
\end{figure*}

Gru~II is the third UFD galaxy in which an enhancement in $r$-process elements has been detected, with Ret~II and Tuc~III being the other two. The stars in Tuc~III and Ret~II exhibit a more uniform enhancement in $r$-process elements, with all five stars analysed in Tuc~III showing a mild $r$-process enhancement ($\mathrm{[Eu/Fe]} \sim 0.5$), and seven of the nine stars in Ret~II being extremely enhanced ($\mathrm{[Eu/Fe]} > 1$). 
In Figure \ref{fig:rproc} the neutron-capture element abundances in J220423 are compared to the scaled solar system $r$-process residuals from \citet{arlandini1999}. The scaling factor is calculated from the average difference between the J220423 and solar abundances for Ba, La, Pr, Nd, and Eu. The neutron-capture abundance pattern in J220423 matches the scaled solar system $r$-process abundance pattern for elements from Ba to Eu. A similar match has also been found for $r$-process enhanced stars in other ultra-faint and classical dwarf galaxies \citep{hansen2017,ji2016c,marshall2019}, suggesting that similar nucleosynthesis events enriched these galaxies. Currently three astrophysical sites for heavy $r$-process element production are proposed; NSM \citep{lattimer1974}, magnetorotational SN (Jet-SN) \citep{winteler2012}, and collapsars \citep{siegel2019}. Observationally only the NSM has been confirmed to produce $r$-process elements \citep{drout2017}. Also, more recent models for the Jet-SNe do not find that this site is capable of producing the heavy $r$-process elements \citep{mosta2018}.

For Ret~II, \citet{ji2016a} argued that the most likely source of $r$-process elements in this galaxy is a NSM, due to the small size of the galaxy and the very large $r$-process element enrichment. It should be noted that collapsars, which could be very efficient $r$-process production sites \citep{siegel2019}, were only introduced as a possible $r$-process element production site after the \citet{ji2016a} study. With the somewhat milder enhancement of Tuc~III it was speculated that Tuc~III may have been more massive in the past, thus having more gas to dilute the ejecta from the $r$-process nucleosynthesis event compared to Ret~II \citep{hansen2017,marshall2019}. This scenario also agrees with the extended tails detected for Tuc~III \citep{li2018}. Alternatively, the gas in Tuc~III could also have been polluted by an outside event and thus only received a fraction of the $r$-process ejecta. 

When exploring the possible sources for the $r$-process elements detected in Gru~II we can consider the possible neutron-capture element production from the progenitors of the $\alpha$-element signature seen in Gru~II. For the stars in Her also exhibiting the high $\mathrm{[Mg/Ca]}$ signature, \citet{koch2008} determined upper limits of $\mathrm{[Ba/Fe]} < -2.1$. In fact, \citet{koch2013} detected a general deficiency of Ba in a sample of 20 stars in Her. Thus the neutron-capture element enhancement of one of the Gru~II stars does not seem to be coupled with the high-mass CCSNe associated with the $\alpha$-element signature detected for some stars in Her and all three stars analysed in Gru~II. In fact, this type of CCSN is likely inefficient in producing neutron-capture elements. 

The possible top-heavy nature of the IMF of Gru~II, as suggested by the $\alpha$-element abundances of the stars, would result in an increased number of SNe. This in principle increases the chance of all three proposed $r$-process nucleosynthesis sites to occur and does not provide any evidence of which one is most likely to have occurred. Following \citet{ji2019b} and using the solar $r$-process abundances from \citet{sneden2008}, we calculate the lanthanide fraction $X_{\rm La}$ for J220423 and find $\log X_{\rm La}=-1.2$. This value is very similar to the lanthanide fractions of Tuc~III ($\log X_{\rm La}=-1.5$) and Ret~II ($\log X_{\rm La}=-1.1$) \citep{ji2019b}, and somewhat higher than the value found for GW170817 of $\log X_{\rm La}=-2.2$. With only one NSM detected to date it is unclear if the lanthanide fraction derived from GW170817 represents the general lanthanide fraction distribution for NSMs; it is therefore difficult to conclude with this number that a NSM is not the source of $r$-process elements in Gru~II. In fact, the fact that J220409 and J220352, the most metal-poor stars of the three, show no Eu enhancement suggest some delay time from the enrichment of the system with $\alpha$ elements to $r$-process element enrichment, pointing at a NSM as the origin of the $r$-process elements in Gru~II. It is also possible that the $r$-process event enriching the gas from which J220423 formed happened in a neighboring system, in which case the lanthanide fraction may be the only clue to the origin of these elements.

\section{Summary \label{sec:sum}}
In this paper we have derived abundances for the three brightest member stars at the top of the giant branch of the UFD galaxy Gru~II. High $\mathrm{[Mg/Ca]}$ ratios were determined for all stars. This abundance signature can be produced via nucleosynthesis in high-mass ($\geqslant 20$M$_\odot$) CCSNe \citep{woosley1995,heger2010}, suggesting that Gru~II was mainly enriched by a population of very high-mass stars and has a top-heavy IMF. The same chemical signature was also detected in the UFD galaxy Her \citep{koch2008}, where the low mass stellar population has been found to have a top-heavy IMF \citep{geha2013}. Alternatively the chemical signature of Gru~II can also be the result of stochastic chemical enrichment of the galaxy. To further explore this issue, abundances for a larger sample of stars in Gru~II covering a wider metallicity range is needed. 

The abundances of Gru~II also revealed an enhancement in $r$-process elements in the most metal-rich of the three stars analysed. This star displays the well-known match to the scaled solar system $r$-process abundance pattern, similar to what has been found previously for other $r$-process enhanced stars in ultra-faint and classical dwarf galaxies and the MW halo \citep{sneden2008,ji2016c,hansen2017}. The progenitor of the $r$-process enhancement of Gru~II does not seem to be directly coupled with the $\alpha$-element signature. However, the possible top-heavy nature of the IMF of Gru~II would result in a larger population of high-mass stars in Gru~II and thus also a higher possibility of various types of SNe, leading to $r$-process element production. We calculate a lanthanide fraction of $\log X_{\rm La}=-1.2$, similar to the fractions found in the other two UFD galaxies with $r$-process enhanced stars, Ret~II and Tuc~III, and higher than the value derived for GW170817 \citep{ji2019b}, thus not directly supporting a NSM origin for the $r$-process elements. The possible delay in the $r$-process enhancement compared to the $\alpha$-enhancement of the galaxy, however, does support a NSM as the source of the $r$-process material in this galaxy. 

The relationship between the IMF of UFDs and their chemical abundances is relatively unexplored, mainly due to the limited measurements of IMFs of UFD galaxies in combination with the small number of stars in these systems for which abundances can be measured. Expanding this data set will help to better understand the chemical evolution of these systems and the nucleosynthesis of the first stars.

\acknowledgements{
The authors thank the referee for a careful reading of the manuscript and A. McWilliam for useful discussion of abundance error propagation.

Funding for the DES Projects has been provided by the U.S. Department of Energy, the U.S. National Science Foundation, the Ministry of Science and Education of Spain, the Science and Technology Facilities Council of the United Kingdom, the Higher Education Funding Council for England, the National Center for Supercomputing Applications at the University of Illinois at Urbana-Champaign, the Kavli Institute of Cosmological Physics at the University of Chicago, 
the Center for Cosmology and Astro-Particle Physics at the Ohio State University, the Mitchell Institute for Fundamental Physics and Astronomy at Texas A\&M University, Financiadora de Estudos e Projetos, Funda{\c c}{\~a}o Carlos Chagas Filho de Amparo {\`a} Pesquisa do Estado do Rio de Janeiro, Conselho Nacional de Desenvolvimento Cient{\'i}fico e Tecnol{\'o}gico and the Minist{\'e}rio da Ci{\^e}ncia, Tecnologia e Inova{\c c}{\~a}o, the Deutsche Forschungsgemeinschaft and the Collaborating Institutions in the Dark Energy Survey. 

The Collaborating Institutions are Argonne National Laboratory, the University of California at Santa Cruz, the University of Cambridge, Centro de Investigaciones Energ{\'e}ticas, Medioambientales y Tecnol{\'o}gicas-Madrid, the University of Chicago, University College London, the DES-Brazil Consortium, the University of Edinburgh, the Eidgen{\"o}ssische Technische Hochschule (ETH) Z{\"u}rich, Fermi National Accelerator Laboratory, the University of Illinois at Urbana-Champaign, the Institut de Ci{\`e}ncies de l'Espai (IEEC/CSIC), the Institut de F{\'i}sica d'Altes Energies, Lawrence Berkeley National Laboratory, the Ludwig-Maximilians Universit{\"a}t M{\"u}nchen and the associated Excellence Cluster Universe, the University of Michigan, the National Optical Astronomy Observatory, the University of Nottingham, The Ohio State University, the University of Pennsylvania, the University of Portsmouth, SLAC National Accelerator Laboratory, Stanford University, the University of Sussex, Texas A\&M University, and the OzDES Membership Consortium.

Based in part on observations at Cerro Tololo Inter-American Observatory, National Optical Astronomy Observatory, which is operated by the Association of 
Universities for Research in Astronomy (AURA) under a cooperative agreement with the National Science Foundation.

The DES data management system is supported by the National Science Foundation under Grant Numbers AST-1138766 and AST-1536171.
The DES participants from Spanish institutions are partially supported by MINECO under grants AYA2015-71825, ESP2015-66861, FPA2015-68048, SEV-2016-0588, SEV-2016-0597, and MDM-2015-0509, some of which include ERDF funds from the European Union. IFAE is partially funded by the CERCA program of the Generalitat de Catalunya.
Research leading to these results has received funding from the European Research Council under the European Union's Seventh Framework Program (FP7/2007-2013) including ERC grant agreements 240672, 291329, and 306478. We  acknowledge support from the Brazilian Instituto Nacional de Ci\^encia
e Tecnologia (INCT) e-Universe (CNPq grant 465376/2014-2).

T.T.H and J.D.S acknowledge support from NSF grant AST-1714873.  TSL is supported by NASA through Hubble Fellowship grant HST-HF2-51439.001, awarded by the Space Telescope Science Institute, which is operated by the Association of Universities for Research in Astronomy, Inc., for NASA, under contract NAS5-26555.

This manuscript has been authored by Fermi Research Alliance, LLC under Contract No. DE-AC02-07CH11359 with the U.S. Department of Energy, Office of Science, Office of High Energy Physics.}

\facilities{Magellan/Clay MIKE}
\software{MOOG \citep{sneden1973,sobeck2011}, IRAF \citep{tody1986,tody1993},  ATLAS9 \citep{castelli2003},
linemake (https://github.com/vmplacco/linemake), STARFIT (http://starfit.org/)}


\end{document}